\documentclass[aps,prl,reprint,superscriptaddress,longbibliography]{revtex4-2}
\usepackage{amsmath,amssymb,bm,graphicx,booktabs}
\usepackage[colorlinks=true,citecolor=blue,linkcolor=blue,urlcolor=blue]{hyperref}
\hypersetup{pdftitle={Fidelity Determines the Minimum Error of Quantum Recovery},
pdfauthor={Dario Picozzi}}
\newcommand{\id}{\mathrm{id}}
\newcommand{\Tr}{\operatorname{Tr}}
\newcommand{\Ad}{\operatorname{Ad}}
\newcommand{\smin}{s_{\min}}
\newcommand{\lmax}{\lambda_{\max}}

\newcommand{\cN}{\mathcal N}
\newcommand{\cR}{\mathcal R}
\newcommand{\cC}{\mathcal C}
\newcommand{\cE}{\mathcal E}
\newcommand{\norm}[1]{\left\lVert#1\right\rVert}
\newcommand{\ket}[1]{\lvert#1\rangle}
\newcommand{\bra}[1]{\langle#1\rvert}
\newcommand{\proj}[1]{\ket{#1}\bra{#1}}
\newcommand{\diag}{\operatorname{diag}}
\newcommand{\suppmat}{Supplemental Material}

\begin{document}
\title{Fidelity Determines the Minimum Error of Quantum Recovery}
\author{Dario Picozzi}
\email{picozzi.dario@gmail.com}
\affiliation{Department of Physics and Astronomy, University College London (UCL),
Gower Street, London WC1E 6BT, United Kingdom}
\affiliation{London Centre for Nanotechnology, 19 Gordon St,
London WC1H 0AH, United Kingdom}

\begin{abstract}
Even a recovery that maximizes worst-input entanglement fidelity can
retain avoidable errors in subsequent measurements. Entanglement fidelity
is the probability of passing a test of whether the state and its
entanglement were preserved; infidelity is the probability of failing it.
Operational error is the largest possible change in any measurement
probability, quantified by the normalized diamond distance. For every
finite-dimensional noise channel and fixed encoding, we prove that
operational error and worst-input entanglement infidelity have
the same minimum over all recovery operations. The proof constructs one
final unitary (a reversible quantum operation) that makes any recovery's
operational error no greater than its original worst-input entanglement
infidelity. With a retained noise record, we determine the optimal error
and characterize all optimal decoders subject to a Kraus-rank limit for
each outcome. For a fixed decoded channel, a semidefinite program finds
the optimal final unitary in the low-error regime. We characterize all
optimal final corrections and show that, at a fixed error floor
$0<\epsilon<1$, the operational optimizers' sensitivity to residual unitary errors of size $s$
is at worst of order $s^{4/3}$, while the canonical polar correction
has quadratic sensitivity.
\end{abstract}
\maketitle

How accurately can a quantum memory preserve an unknown state and its
entanglement with a reference? Entanglement fidelity is the probability
of passing a test that the joint state was preserved; infidelity is the
probability of failing it. Operational error is the largest change in any
measurement-outcome probability, optimized over inputs and joint
measurements on the system and reference. Even a recovery that maximizes
worst-input entanglement fidelity can retain avoidable operational error.
We show that one final unitary can make every such recovery operationally
optimal.

Entanglement fidelity quantifies preservation of quantum
correlations~\cite{Schumacher1996}. For approximate error correction,
B\'eny and Oreshkov characterized its worst-input recovery optimum through
a dual problem on the environment~\cite{BenyOreshkov2010,BenyOreshkov2011}.
The diamond distance supplies a worst-case operational error that remains
valid when the system is entangled with a reference. General fidelity--distance
comparisons permit a square-root gap~\cite{FuchsGraaf1999,Kretschmann2008}.
Coherent phase errors can produce such square-root scaling~\cite{Kueng2016}.
Does optimizing a recovery remove this gap?
Holdsworth, Singh and Wilde proved equality of the optimized fidelity and
diamond errors when encoder and decoder can coordinate their actions
using free shared randomness. They explicitly left the equality without
that assistance open~\cite{Holdsworth2023}.

Here we prove this equality for every finite-dimensional noise channel
and every fixed encoding. The construction acts only on the output
of a supplied decoder: one unitary makes its operational error no larger
than its original worst-input entanglement infidelity. Fidelity optimization
therefore determines the minimum operational error; the final unitary ensures
that the chosen fidelity-optimal decoder attains it. An additive fidelity
deficit transfers with the same coefficient. The same
principle applies to restricted decoder families closed under output
unitaries and to calibration of a fixed noisy gate.

\begin{samepage}
\emph{Error measures and the recovery principle.}
Let $\Phi$ describe the physical process on a $d$-dimensional logical
system, including any existing recovery. It is a completely positive,
trace-preserving (CPTP) map with Kraus operators $A_j$ giving the matrix
representation $\Phi(X)=\sum_{j=1}^r A_jXA_j^\dagger$, where $X$ is any
operator on the logical system. Define
\begin{equation}
 \begin{aligned}
 F(\Phi)&=\min_\rho\sum_j|\Tr(\rho A_j)|^2,\\
 D(\Phi)&=\tfrac12\norm{\Phi-\id}_\diamond .
 \end{aligned}
 \label{eq:measures}
\end{equation}
\end{samepage}
The minimum is over all logical states $\rho$, represented by density
matrices. A purification $\ket{\psi_\rho}$ is a joint pure state of the
logical system and an untouched reference, with the logical system alone
described by $\rho$.
The sum in Eq.~\eqref{eq:measures} is the probability of returning to
this joint input,
$\bra{\psi_\rho}(\Phi\otimes\id)(\proj{\psi_\rho})\ket{\psi_\rho}$.
We use the squared-fidelity convention for entanglement
fidelity~\cite{Schumacher1996}. For example, one half of the Bell pair
$(\ket{00}+\ket{11})/\sqrt2$ has marginal $I/2$. Sending that half through
the memory while leaving its reference untouched tests whether their
joint state survives. Minimizing over $\rho$ includes all such joint inputs.
The normalized distance $D$ is the
largest change in a measurement-outcome probability, optimizing over
inputs, reference systems and measurements~\cite{Watrous2018}.
Thus $D=0.01$ bounds the change in every such probability by one
percentage point. Testing the projector onto
the ideal purified output gives $D(\Phi)\ge1-F(\Phi)$; the general upper
bound is $D(\Phi)\le\sqrt{1-F(\Phi)}$.
For a unitary $U$, representing a reversible operation, write
$\Ad_U(X)=UXU^\dagger$ for its channel.

\textbf{Unitary recovery principle.}
For every channel $\Phi$ there is an output unitary $U$ satisfying
the sharp, dimension-independent bound
\begin{equation}
 1-F(\Ad_U\Phi)\le D(\Ad_U\Phi)\le1-F(\Phi)\le D(\Phi).
 \label{eq:principle}
\end{equation}
Both fidelity and operational error improve or remain unchanged. The correction is
calculated from the characterized channel and acts only on the logical output.

Let $\cN$ describe the physical noise together with a fixed encoding,
and let $\cR$ denote the decoder, a CPTP map from the noisy physical
output to the logical system.
Products of maps denote composition.
Consider a nonempty decoder family $\cC$ in which appending any final
logical unitary keeps a recovery in the family:
$\Ad_U\cR\in\cC$ for every $\cR\in\cC$ and logical unitary $U$.
Applying Eq.~\eqref{eq:principle} to $\Phi=\cR\cN$ gives
\begin{equation}
 \inf_{\cR\in\cC}D(\cR\cN)
        =1-\sup_{\cR\in\cC}F(\cR\cN).
 \label{eq:recovery}
\end{equation}
The lower bound $D\ge1-F$ gives one direction; applying the correction
to decoders approaching the fidelity supremum gives the other.
The argument covers nonconvex families and unattained optima.
The unrestricted CPTP optima are attained; optimizing the encoder afterwards also preserves
the equality. This resolves the unassisted question of
Ref.~\cite{Holdsworth2023} with a deterministic output correction.

The same correction preserves additive accuracy: a fidelity deficit of
at most $\eta\ge0$ gives an operational error excess of at most $\eta$,
relative to the respective optimal values in $\cC$.
Output-unitary correction preserves the decoder's Kraus rank, so fixed
rank caps are included. Every diamond optimizer is fidelity optimal, since
$D\ge1-F$ and Eq.~\eqref{eq:recovery} force equality at the optimum;
a fidelity optimizer can still need the correction. A short two-qubit
counterexample in the \suppmat~\cite{Supplement} establishes this
strict distinction between the optimizer sets.

\emph{Construction and proof.}
The worst-preserved input selects a coherent combination of noise
amplitudes. Polar decomposition separates this combination into a
rotation and a positive filter. The correction cancels the rotation;
the proof bounds the remaining operational error by the original infidelity.
Choose a minimizing state $\rho_0$ in Eq.~\eqref{eq:measures} and take the
polar decomposition
\begin{equation}
 \begin{split}
 f&=F(\Phi),\qquad \alpha_j=\Tr(\rho_0 A_j),\\
 T&=\sum_j\bar\alpha_j A_j=U^\dagger|T|.
 \end{split}
 \label{eq:construction}
\end{equation}
Here $|T|=(T^\dagger T)^{1/2}$ and
$\operatorname{Re}T=(T+T^\dagger)/2$ is the Hermitian part.
At $f=0$ the guarantee is the universal bound $D\le1$. For $f>0$,
first-order optimality along the line from $\rho_0$ to any state $\sigma$
gives $\Tr(\sigma\operatorname{Re}T)\ge f$. Thus
$\operatorname{Re}T\succeq fI$, meaning that every eigenvalue is at least $f$.
For any unit vector $v$,
$\norm{Tv}\ge\operatorname{Re}\langle v,Tv\rangle\ge f$, so
$|T|\succeq fI$ as well.
This recovers the known weighted Kraus inequality~\cite{FungChau2014}.
It also makes $T$ invertible, so the correction $U$ is unique.
Tied minimizing states yield the same weighted
amplitude $T$ (End Matter).

Polar decomposition of selected Kraus operators separates coherent
and decoherent errors~\cite{Carignan2019}.
After polar correction of a selected Kraus operator $A$, an earlier bound
gives operational error at most $1-\smin(A)^2$, provided
$s_{\max}(A)\le3\smin(A)$~\cite{SkanesNorman2021}.
Here $\smin$ and $s_{\max}$ are the smallest and largest singular values.
The centered identity below removes this restriction. Combined with
the worst-input amplitude, it gives the sharp recovery principle
and equality of optimized errors.

Set $L_j=UA_j-\alpha_j I$. Trace preservation and $UT=|T|$ give
\begin{equation}
 \begin{gathered}
 H=\sum_jL_j^\dagger L_j=(1+f)I-2|T|,\\
 0\preceq H\preceq(1-f)I .
 \end{gathered}
 \label{eq:gram}
\end{equation}
Exact expansion yields
\begin{equation}
 (\Ad_U\Phi-\id)(X)=\sum_jL_jXL_j^\dagger-\tfrac12(HX+XH).
 \label{eq:centered}
\end{equation}
For a completely positive (CP) map $\Lambda$, its diamond norm is
$\norm{\Lambda^\dagger(I)}_\infty$~\cite{Watrous2018}, where
$\norm{\cdot}_\infty$ is the operator norm. The CP term therefore has
diamond norm $\norm{H}_\infty$. Left and right multiplication by $H$
each have norm at most $\norm{H}_\infty$, so the anticommutator term
obeys the same bound. Dividing their sum by two
gives $D(\Ad_U\Phi)\le\norm{H}_\infty\le1-f$, proving
Eq.~\eqref{eq:principle}. The centered identity holds exactly at every
noise strength.
Qubit dephasing, $(1-p)\id+p\Ad_Z$ with $0\le p\le\tfrac12$, saturates the
bound: $F=1-p$ and $D=p$.

For $f>0$, the corrected channel also has a transparent structure.
A unitary change of Kraus coordinates makes one Kraus operator
$K=|T|/\sqrt f\succeq\sqrt f I$. It is positive and contractive, and
\begin{equation}
 \begin{split}
 (\Ad_U\Phi)(X)&=KXK+\cE(X),\\
 \cE^\dagger(I)&=I-K^2\preceq(1-f)I,
 \end{split}
 \label{eq:positive}
\end{equation}
where $\cE$ is CP. The filter $K$ attenuates amplitudes in its eigenbasis
without adding phases. In this Kraus representation, its probability
is at least $f$ on every input; the remaining terms have total probability
at most $1-f$.
\par

\begin{figure*}[t]
 \includegraphics[width=\textwidth]{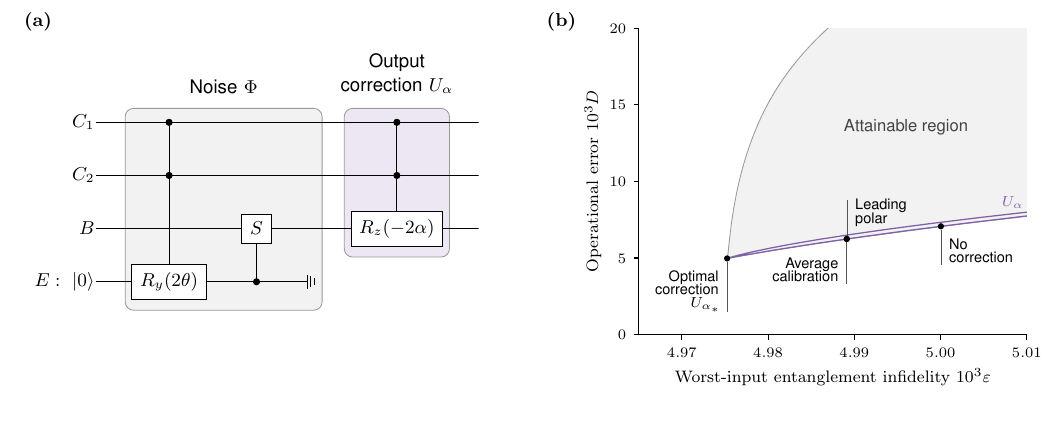}
 \caption{Three-qubit conditional-gate calibration.
 (a) Conditional noise $\Phi$, followed by output correction $U_\alpha$.
 On the control branch $\ket{11}$, a flag occurs with probability
 $p=\sin^2\theta$ and triggers $S=\diag(1,i)$ on $B$.
 The environment $E$ is then discarded; $U_\alpha$ acts only on the data.
 (b) Worst-input entanglement infidelity $\varepsilon=1-F$ versus
 operational error $D$ at $p=0.01$, for ideal corrections.
 The gray region shows the pairs attainable by arbitrary unitary corrections;
 the purple curve traces the family $U_\alpha$ in (a).
 The region's vertical extent shows how much operational error can vary
 at fixed infidelity.
 Markers show no correction ($\alpha=0$), leading-Kraus polar
 correction, calibration maximizing average gate fidelity, and the optimal
 correction $U_{\alpha_*}$ in Eq.~\eqref{eq:calibration}.
 The correction $U_{\alpha_*}$ minimizes both plotted errors; the average
 gate fidelity and operational error optima are over all CPTP recoveries.
 Operational error uses $D(\Phi)=\norm{\Phi-\id}_\diamond/2$.
 }
 \label{fig:examples}
\end{figure*}

\emph{Recovery with a retained record.}
Let $A_{si}:\mathbb C^d\to\mathbb C^{m_s}$ be the noise Kraus operators
for observed outcome $s$ and hidden index $i$, with
$\sum_{s,i}A_{si}^\dagger A_{si}=I$.
Write $F_{\boldsymbol k}^*$ and $D_{\boldsymbol k}^*$ for the optimal values
with at most $k_s$ Kraus operators in the trace-preserving decoder for $s$,
where $k_sd\ge m_s$. Let $\mathcal T_{\boldsymbol k}$ consist of positive
semidefinite block matrices $\tau=\bigoplus_s\tau_s$ of trace one with
$\operatorname{rank}\tau_s\le k_s$. Then
\begin{equation}
 \begin{aligned}
 F_{\boldsymbol k}^*=1-D_{\boldsymbol k}^*
  &=\max_{\tau\in\mathcal T_{\boldsymbol k}}\min_\rho
    \left[\sum_s\Tr\sqrt{G_s(\rho,\tau_s)}\right]^2,\\
 G_s(\rho,\tau_s)
  &=\sum_{i,j}(\tau_s)_{ji}A_{si}\rho^2 A_{sj}^\dagger .
 \end{aligned}
 \label{eq:recorded-optimum}
\end{equation}
The minimum uses one logical density matrix $\rho$ for all outcomes.
The environment formula builds on rank-constrained fidelity
duality~\cite{BenyOreshkov2011}. A polar construction attains both optima
with one deterministic decoder at the same budgets, including for singular
sources (\suppmat). When all branches are square, conditional unitaries suffice
precisely when an unrestricted maximizing reference has rank at most
one in each block. A single square branch with $k_1=1$ gives the
final-unitary problem below.
\par
\emph{Computing the best final unitary.}
A fixed decoded channel can admit a better correction than the guaranteed
choice in Eq.~\eqref{eq:construction}. The same polar construction yields
its globally best output unitary when the coherent Kraus weights are
optimized. Whenever the best normalized error is below one half,
a single semidefinite program (SDP) determines those weights exactly.
Consider an $r$-dimensional density matrix $\tau$ and the program
\begin{equation}
 \begin{split}
 \Gamma(\Phi)=\max_{\tau,t}\ &t\\
 \text{subject to }&\tau\succeq0,\quad\Tr\tau=1,\\
 &\sum_{ij}\tau_{ij}A_i^\dagger A_j\succeq tI .
 \end{split}
 \label{eq:sdp}
\end{equation}

The positive matrix $\tau$ replaces a single vector of amplitude weights,
making the optimization convex.
If $\Gamma>\tfrac12$, every optimal $\tau$ is rank one, so it determines one
coefficient vector and hence a unitary correction. Write $\tau=cc^\dagger$,
form $B_c=\sum_j\bar c_jA_j=U_c^\dagger|B_c|$, and apply
$U_c$. Among all final unitaries, this correction attains
the maximum fidelity $\Gamma$ and the minimum diamond error $1-\Gamma$,
certifying the error floor for this calibration.
The rank-one proof and dual certificate are in End Matter. The test
$\Gamma>\tfrac12$ certifies exactness directly from the SDP output. The SDP uses
$d\times d$ Kraus matrices and an $r\times r$ density matrix;
one may choose $r\le d^2$.
Beyond this regime, a convergent hierarchy finds near-optimal corrections
with certified error bounds (\suppmat).
\par
\emph{Selecting an optimal correction.}
Equal optimal values do not imply equal sets of optimizers.
For a fixed decoded channel, write
$q=\max_U F(\Ad_U\Phi)$ and $\epsilon=1-q$, with $0<q<1$.
Each fidelity-optimal gate determines an optimal coherent amplitude
$B_c=\sum_j\bar c_jA_j$ and its polar correction $U_c$.

For each fixed $B_c$, the associated optimal gates differ from $U_c$
by residual rotations. These rotations fix the eigenspace of $|B_c|$
at its smallest eigenvalue $\sqrt q$. In Hermitian coordinates for the
rotations (the Cayley transform), the two complete optimal families
form nested convex sets, with the operational family inside the
fidelity family (End Matter). In the physical gate distance
$d(U,V)=\norm{\Ad_U-\Ad_V}_\diamond/2$, they obey the sharp bounds
\begin{equation}
 \begin{split}
 F(\Ad_U\Phi)=q&\ \Longrightarrow\ d(U,U_c)\le\sqrt\epsilon,\\
 D(\Ad_U\Phi)=\epsilon&\ \Longrightarrow\
 d(U,U_c)\le\tfrac\epsilon2\sqrt{3+\epsilon}.
 \end{split}
 \label{eq:optimal-radii}
\end{equation}
The sharp enclosing radii therefore scale as $\sqrt\epsilon$ for
fidelity optimality and as $\epsilon$ for operational optimality.
When $q>\tfrac12$, the SDP fixes a unique coherent amplitude up to phase.
If $|B_c|=\sqrt q I$, both families collapse to the same gate, up to phase.
Otherwise the operational family is strictly smaller;
both contain continuously many gates.

Among these nominally optimal choices, the polar correction has a
stronger universal guarantee against residual unitary error.
Let a residual unitary $V$ of size $s=d(V,I)$ act after the corrected
channel. For the canonical polar correction, $\Lambda=\Ad_{U_c}\Phi$,
the sharp asymptotic bound is
\begin{equation}
 D(\Ad_V\Lambda)-\epsilon
 \le \frac{\sqrt q(1+\sqrt q)^2}{(2+\sqrt q)\epsilon}s^2
       +O_q(s^4).
 \label{eq:polar-robustness}
\end{equation}
For an arbitrary operationally optimal gate,
the universal bound is instead $O_q(s^{4/3})$, and the exponent is
sharp. A three-level loss channel has two nominally optimal gates
whose excess errors under the same physical drift scale as $s^2$
and $s^{4/3}$, respectively. The proof uses the exact coherent-amplitude
certificate in End Matter; the \suppmat\ gives the examples.
At zero noise the error is linear in $s$. The polar construction therefore
provides a quadratic robustness guarantee in addition to attaining
the nominal optimum.
\par
\emph{Three-qubit conditional-gate calibration.}
Consider two control qubits $C_1,C_2$ and a target $B$.
Only the control branch $\ket{11}$ couples to an environment flag,
initially in $\ket0$. A flag occurs with probability $p$ on that branch
and applies $S=\diag(1,i)$ to $B$ [Fig.~\ref{fig:examples}(a)].
Discarding the flag gives the error channel
\begin{equation}
 \begin{split}
 A_0&=[I_C-(1-\sqrt{1-p})Q]\otimes I_B,\\
 A_1&=\sqrt p\,Q\otimes S,\qquad Q=\proj{11}_{C_1C_2}.
 \end{split}
 \label{eq:model}
\end{equation}
All eight data levels carry logical information. The two active levels
are $\ket{110}$ and $\ket{111}$; the remaining six are quiet.
Corrections access only the data output.

\begin{samepage}
The optimal recovery is a doubly controlled target rotation,
\begin{equation}
 \begin{split}
 U_\alpha&=e^{i\alpha Q\otimes Z_B},\qquad
 \alpha_*=\tfrac12\arcsin p,\\
 D_*&:=\min_{\cR\ {\rm CPTP}}D(\cR\Phi)
   =\frac{1-\sqrt{(1-p)/(1+p)}}2 .
 \end{split}
 \label{eq:calibration}
\end{equation}
\end{samepage}
This unitary attains the optimum over all CPTP recoveries for
$0\le p\le1$. For $0<p<1$, the unique recovery maximizing average gate
fidelity~\cite{Nielsen2002} is another member $U_{\alpha_{\rm av}}$ of
the same family. At weak noise,
$\alpha_{\rm av}\sim p/8$, whereas $\alpha_*\sim p/2$.
Average gate fidelity averages the return probability over uniformly
distributed pure inputs. Maximizing it is equivalent to maximizing
entanglement fidelity at the maximally mixed logical state $I/8$, which
gives the active branch only one quarter of the weight. A worst input of
the optimally corrected channel has active population $1/(1+p)$,
approaching one at weak noise. The average criterion therefore
undercorrects the branch that a computation can occupy almost entirely.
End Matter proves the recovery optimum; the \suppmat~\cite{Supplement}
proves the average comparator's uniqueness.

At $p=0.01$, diamond errors are $0.007071$ without correction,
$0.006246$ for prior leading-Kraus polar correction~\cite{Carignan2019},
$0.006244$ for the globally average-optimal recovery, and
$0.004975$ for Eq.~\eqref{eq:calibration}. The last is a
$20.3\%$ reduction beyond the average optimum.
Figure~\ref{fig:examples}(b) shows that unitary corrections can have the
same worst-input fidelity but substantially different operational errors.

\looseness=-1
The same example directly illustrates the recovery principle.
The average-calibrated channel already has worst-input fidelity $0.99501089$,
within $1.39\times10^{-5}$ of the unrestricted optimum $0.99502475$.
By Eq.~\eqref{eq:principle}, this fidelity guarantees a corrected error
no larger than its original infidelity, $0.00498911$.
The construction in Eq.~\eqref{eq:construction} supplies that correction.
The actual repaired error is approximately $0.00497526$, almost the
exact optimum. Thus a small fidelity deficit can conceal an appreciable
operational error, and the principal construction removes it.

The distinction is visible in a simple local measurement. Prepare
$\ket{11+}$ and measure $Y$ on $B$, whose ideal expectation is zero.
The target coherence is multiplied by
$z_\alpha=(1-p-ip)e^{2i\alpha}$, so $\langle Y\rangle=-\operatorname{Im}z_\alpha$.
Average calibration gives $0.00751$, while optimal correction gives
$0.0000995$.

All three corrections use the same four-CNOT, four-$R_z$ circuit.
In the implementation-noise model specified in the \suppmat, the
correction retains a $12.0\%$ advantage over average calibration
reoptimized within that circuit family.

The optimal coherent amplitude here has constant singular values.
This gives a finite calibration tolerance: an additional coherent gate
error of size $s$ increases the error floor $\epsilon=D_*$ by at most
$(\sqrt{\epsilon^2+4(1-\epsilon)s^2}-\epsilon)/2$
(\suppmat). At $p=0.01$, $s\le0.00165$ therefore keeps the
operational error within $10\%$ of its optimum.
\par
For a fixed quantum code, correction turns a bound on worst-input
entanglement infidelity into an equally small bound on changes in any
measurement probability, including with an entangled reference.
Fidelity constructions and converses therefore bound optimal operational
error with the same constants. For example, correcting the approximate
recovery of B\'eny and Oreshkov~\cite{BenyOreshkov2011} gives diamond error
at most four times the optimum over all CPTP recoveries (\suppmat).

For a channel with equal input and output dimensions and no retained
noise record, let $e$ be the minimum error over all CPTP output corrections
and $u$ the minimum over unitary corrections of that same channel.
Then $0\le u-e\le e^2/(1-e)$, with sharp leading coefficient one as
$e\to0$ (\suppmat). When the calibration SDP finds $u<1/2$, it therefore
certifies that $e$ lies between $u/(1+u)$ and $u$.

For a circuit of noisy gates with unitary targets, these guarantees
compose: the total error after correction is bounded by the sum of
the gates' original worst-input entanglement infidelities.
Channel and implementation error bounds add to this budget
(\suppmat~\cite{Supplement}).

Independent parallel blocks admit a stronger conclusion. If each block
has optimal final-unitary error $\epsilon_i<\tfrac12$, the product of its
locally computed polar corrections is globally optimal, even among
entangling final unitaries:
\begin{equation}
 \min_{U\ {\rm joint}}D\!\left(\Ad_U\bigotimes_i\Phi_i\right)
       =1-\prod_i(1-\epsilon_i).
 \label{eq:local-global}
\end{equation}
This includes inputs entangled across all blocks and with a reference.
Only the local errors must be below one half; the combined error need not
be. The positive SDP product argument~\cite{LeeMittal2008} and the
local rank-one certificates prove the result (End Matter).

The recovery principle also extends to pure isometric target operations in larger
output spaces. The supplement gives this extension, finite certificates
and a standard-code calibration control. Optimized infidelity and
operational error coincide; one final unitary connects their optimizers.
\par

\begin{acknowledgments}
The author acknowledges support from the Engineering and Physical Sciences
Research Council (EPSRC) under grants EP/T517793/1, EP/S021582/1 and
EP/W524335/1. Generative AI assisted with research and manuscript
preparation. The author directed the work, verified the outputs and
takes responsibility for the content.
\end{acknowledgments}
\paragraph*{Data and code availability.}
Data and code supporting the findings are available from the
corresponding author on reasonable request.

\nocite{Yu2022,Gottesman1996,Hao2021}
\bibliography{references}
\clearpage
\section*{End Matter}
\appendix
\brokenpenalty=10000
\section{Independence of the minimizing state}
At a worst input the vector $\alpha_j=\Tr(\rho_0A_j)$ has squared norm
$f$. If two minimizing states gave different vectors $\alpha,\beta$,
their midpoint would have score
$\norm{(\alpha+\beta)/2}_2^2=f-\norm{\alpha-\beta}_2^2/4<f$.
Hence the amplitude vector, and therefore $T$, is independent of the
chosen minimizer. An isometric change of Kraus coordinates also leaves
$T$ invariant.

\section{Optimal final-unitary correction}
\label{sec:unitary-proof}
For a normalized coefficient vector $c$, let
$B_c=\sum_j\bar c_jA_j=U_c^\dagger|B_c|$ and
$q(c)=\smin(B_c)^2$. A Kraus-coordinate change selects $B_c$ as one
Kraus operator. The polar correction $U_c$ makes it positive.
If a channel has a positive Kraus operator $K\succeq\mu I$ with $\mu\ge0$,
center that operator by $\mu I$ and leave the others unchanged. The centered Gram
matrix is $(1+\mu^2)I-2\mu K\preceq(1-\mu^2)I$.
The proof of Eq.~\eqref{eq:centered} then gives $D\le1-\mu^2$,
and its fidelity is at least $\mu^2$. Hence each $c$ gives
$D(\Ad_{U_c}\Phi)\le1-q(c)$ and
$F(\Ad_{U_c}\Phi)\ge q(c)$.
Conversely, applying Eq.~\eqref{eq:construction} to any $\Ad_U\Phi$
of positive fidelity gives a normalized $c$ with $q(c)\ge F(\Ad_U\Phi)$;
at zero fidelity this inequality is automatic. Therefore
\begin{equation}
 q_*:=\max_{\norm{c}_2=1}q(c)
    =\max_U F(\Ad_U\Phi)=1-\min_U D(\Ad_U\Phi).
 \label{eq:variational}
\end{equation}
Singular polar factors may be completed to unitaries. For $f>0$, the guaranteed
weights $c_j=\alpha_j/\sqrt f$ need not maximize $q(c)$ for an arbitrary
starting decoder; they already suffice at a fidelity-optimal decoder
in a unitary-closed family.

For the normalized weighted amplitude, $\smin(B_c)\ge\sqrt f$ and
$s_{\max}(B_c)\le1$. Thus the earlier singular-value condition
already holds for $f\ge1/9$; the centered argument removes the
remaining restriction on noise strength.

Define the complementary channel and its unital adjoint by
\begin{equation}
 \begin{split}
 \widehat\Phi(\rho)_{ij}&=\Tr(A_i\rho A_j^\dagger),\\
 L(\tau)&=\sum_{ij}\tau_{ij}A_i^\dagger A_j.
 \end{split}
 \label{eq:complement}
\end{equation}
The SDP in Eq.~\eqref{eq:sdp} relaxes $\tau=cc^\dagger$ in
Eq.~\eqref{eq:variational}, so $q_*\le\Gamma$. Its dual is
$\Gamma=\min_\rho\lmax(\widehat\Phi(\rho))$, with attained optima
by strict feasibility. For a feasible $L(\tau)\succeq tI$ with
$t>\tfrac12$, put $\lambda=\lmax(\tau)$ and let $c$ be a leading eigenvector.
Positivity and unitality give $\lambda\ge t$. Applying $L$ to the
spectral bound in the first line below gives the second line, because
$L(cc^\dagger)=B_c^\dagger B_c$:
\begin{equation}
 \begin{split}
 \tau&\preceq(1-\lambda)I+(2\lambda-1)cc^\dagger,\\
 tI&\preceq(1-\lambda)I+(2\lambda-1)B_c^\dagger B_c,\\
 q(c)&\ge\frac{t+\lambda-1}{2\lambda-1}
 =t+\frac{(1-\lambda)(2t-1)}{2\lambda-1}.
 \end{split}
 \label{eq:rounding}
\end{equation}
At an optimum $t=\Gamma>\tfrac12$, any $\lambda<1$ would strictly improve
the value. Thus every optimal $\tau$ is rank one and $q_*=\Gamma$.
The threshold is sharp. For qubit reset,
$A_0=\ket0\bra0$, $A_1=\ket0\bra1$, the complementary channel is the
identity and $\Gamma=\tfrac12$. Every $B_c$ has rank one, giving $q_*=0$
and best error one.

\section{Optimal families and their stability}
Put $t=\sqrt q$ and $\epsilon=1-q$, with $0<q<1$.
The following exact certificate is useful independently of optimization:
for any channel $\Psi(X)=\sum_j C_jXC_j^\dagger$,
\begin{equation}
 D(\Psi)=\inf_{\norm{\gamma}_2<1}
 \frac{\norm{I-\sum_j\gamma_j C_j}^2}{1-\norm{\gamma}_2^2}.
 \label{eq:coherent-diamond}
\end{equation}
The infimum is attained whenever $D(\Psi)>0$. For an input marginal
$\rho$, form $G_{ij}=\Tr(\rho C_i^\dagger C_j)$ and
$v_i=\Tr(\rho C_i^\dagger)$. Maximizing $a$ over density matrices $\rho$ subject to
$\left(\begin{smallmatrix}G+aI&v\\v^\dagger&1-a\end{smallmatrix}\right)
\succeq0$ computes $D(\Psi)$.
At a positive optimum this matrix has a one-dimensional kernel;
complementarity forces a rank-one dual optimizer. Factoring that dual
gives Eq.~\eqref{eq:coherent-diamond} (\suppmat).

Fix an optimal normalized coefficient $c$ and write $B_c=PS$, with
$tI\preceq S\preceq I$, and $U=WP^\dagger$. With consistent phase
representatives, the two families are exactly
\begin{equation}
 \begin{split}
 U\text{ fidelity optimal}:&\quad \operatorname{Re}(WS)\succeq tI,\\
 U\text{ operationally optimal}:&\quad
 \operatorname{Re}(WS)\succeq \tfrac t2[(2-q)I+S^2].
 \end{split}
 \label{eq:optimal-families}
\end{equation}
Every optimizer belongs to one such family. The first condition follows
from fidelity stationarity. For the second, write an optimal coefficient
in Eq.~\eqref{eq:coherent-diamond} as $r\bar c$, and put
$b=\smin(B_c)$. The lower bound
$(1-rb)^2/(1-r^2)=1-b^2+(r-b)^2/(1-r^2)\ge1-q$
forces $r=b=t$. Expanding $\norm{I-tUB_c}\le\epsilon$ proves
the second condition. Its right side dominates $t(3-q)S/2$,
while the first dominates $tS$. Testing on eigenvectors of $W$
bounds its eigenphases and proves Eq.~\eqref{eq:optimal-radii}.
The Cayley parametrization $W=(I+iH)(I-iH)^{-1}$ turns each condition
into a convex matrix inequality. It also gives the coincidence and
uniqueness criterion stated in the text (\suppmat).

For an operational optimizer, let $\Lambda$ denote its corrected channel.
Its coherent amplitude $B$ satisfies
$(B-tI)^\dagger(B-tI)\preceq\epsilon(B^\dagger B-qI)$.
Choose the phase of $V$ so $e=\norm{V-I}$ is minimal, and use the
coefficient $r=t-\eta$ in Eq.~\eqref{eq:coherent-diamond}.
The resulting radial margin absorbs the drift cross term by Young's
inequality, giving, for $0<\eta\le t$,
\begin{equation}
 D(\Ad_V\Lambda)-\epsilon
 \le\frac{\eta^2+rt e^2+r\epsilon e^2/\eta}{1-r^2}.
 \label{eq:radial-robustness}
\end{equation}
Balancing $\eta^2/\epsilon$ against $te^2/\eta$ proves
the $O_q(e^{4/3})$ bound; $e=s+O(s^3)$.
For the polar gate, $B=S$ is positive, and optimizing the same
certificate gives Eq.~\eqref{eq:polar-robustness}.

\section{Independent parallel calibration}
Let $L_i=\widehat\Phi_i^\dagger$. Products of local primal and dual
optimizers satisfy
\begin{equation}
 \begin{split}
 (\bigotimes_i L_i)(\bigotimes_i\tau_i)&\succeq(\prod_i\Gamma_i)I,\\
 (\bigotimes_i\widehat\Phi_i)(\bigotimes_i\rho_i)&\preceq(\prod_i\Gamma_i)I.
 \end{split}
\end{equation}
Thus the joint SDP value is $\prod_i\Gamma_i$. When every local
relaxation is tight, the product coherent amplitude has squared least
singular value $\prod_iq_i=\prod_i\Gamma_i$ and attains that bound.
Equation~\eqref{eq:variational} proves Eq.~\eqref{eq:local-global};
$q_i>\tfrac12$ suffices for local tightness.

\section{Exact three-qubit calibration certificate}
\label{sec:examples-proof}
\begin{samepage}
For the channel in Eq.~\eqref{eq:model}, put $s=\sqrt{(1-p)/(1+p)}$, $q=(1+s)/2$ and
$c=(\sqrt q,e^{i\pi/4}\sqrt{1-q})$. The quiet entries of $B_c$ have
squared modulus $q$, and both active entries have squared modulus
\begin{equation}
 (1-p)q+p(1-q)+\sqrt{2p(1-p)q(1-q)}=q .
 \label{eq:calibration-amplitude}
\end{equation}
\end{samepage}
Here $2(1+p)q(1-q)=p$. Thus $B_c$ is $\sqrt q$ times a unitary,
and its inverse polar factor is $U_{\alpha_*}$ in
Eq.~\eqref{eq:calibration}. With $r=\sqrt{1-p^2}$, a Kraus change gives
\begin{equation}
 \begin{split}
 \Psi:=\Ad_{U_{\alpha_*}}\Phi&=q\,\id+(1-q)\Ad_V,\\
 V&=\diag(-I_6,p-ir,p+ir).
 \end{split}
 \label{eq:calibration-mixture}
\end{equation}
The marginal $\rho_p=\diag((p/6)I_6,\tfrac12,\tfrac12)/(1+p)$ commutes with $V$
and obeys $\Tr(\rho_pV)=0$. Thus $\sqrt{\rho_p}$ and
$V^\dagger\sqrt{\rho_p}$ are orthonormal in the Hilbert--Schmidt
inner product. For recovery Kraus operators $R_l$, Bessel's inequality
applied to $R_l\sqrt{\rho_p}$ gives
\begin{equation}
 |\Tr(\rho_pR_l)|^2+|\Tr(\rho_pR_lV)|^2
 \le\Tr(\rho_pR_l^\dagger R_l).
\end{equation}
Since $q\ge\tfrac12$ and $\sum_lR_l^\dagger R_l=I$, the entanglement
fidelity on $\rho_p$ is at most $q$ for every recovery of $\Psi$.
Equation~\eqref{eq:calibration-mixture} attains this bound and diamond
error $1-q$, proving unrestricted optimality.

The product input
\begin{equation}
 \begin{split}
 \ket{\psi_p}&=\ket{c_p}_{C_1}\ket{c_p}_{C_2}\ket+_B,\\
 \ket{c_p}&=\sqrt{1-t}\ket0+\sqrt t\ket1,\qquad
 t=(1+p)^{-1/2}
 \end{split}
 \label{eq:calibration-input}
\end{equation}
has active population $t^2=1/(1+p)$ and hence
$\bra{\psi_p}V\ket{\psi_p}=0$. Its corrected return probability is
$q$; three local return projections attain diamond error $1-q$.
The \suppmat\ proves uniqueness of the average-fidelity comparator
and derives the supplied decoder's worst-input fidelity used in the main text.

\end{document}

\title{Supplemental Material: Fidelity Determines the Minimum Error of Quantum Recovery}
\author{Dario Picozzi}
\email{picozzi.dario@gmail.com}
\affiliation{Department of Physics and Astronomy, University College London (UCL),
Gower Street, London WC1E 6BT, United Kingdom}
\affiliation{London Centre for Nanotechnology, 19 Gordon St,
London WC1H 0AH, United Kingdom}

\maketitle
\setcounter{secnumdepth}{2}
\renewcommand{\theequation}{S\arabic{equation}}
\setcounter{equation}{0}
\renewcommand{\thefigure}{S\arabic{figure}}

We use worst-input entanglement fidelity $F$ in the squared convention
and normalized diamond error $D(\Phi)=\norm{\Phi-\id}_\diamond/2$.
All channels and reference systems are finite dimensional.
Matrix norms without subscripts are operator norms. For unitaries $U,V$,
write $d(U,V)=\norm{\Ad_U-\Ad_V}_\diamond/2$.

Section~\ref{sec:sm-recovery} develops consequences of the recovery principle
and distinguishes fidelity from operational optimizers.
Sections~\ref{sec:sm-coherent}--\ref{sec:sm-critical} prove the optimizer
classification and robustness bounds; Secs.~\ref{sec:sm-precision}
and~\ref{sec:sm-hierarchy} give finite-precision and all-error certificates.
Sections~\ref{sec:sm-calibration} and~\ref{sec:sm-region} develop the
three-qubit example and certify Fig.~1(b).
Section~\ref{sec:sm-recorded} gives the optimal error and classifies the
optimizers for recovery with retained records and prescribed Kraus budgets.
Section~\ref{sec:sm-code} applies the construction to a stabilizer code,
and Sec.~\ref{sec:sm-isometric} extends the principle to isometric targets.

\section{Operational guarantees for quantum recovery}
\label{sec:sm-recovery}

\subsection{Environmental criterion}
Let $\cN$ include the fixed encoding and let $\widehat{\cN}$ be a
complementary channel, describing the information entering the
environment. Write
$D_*:=\min_{\cR\ {\rm CPTP}}D(\cR\cN)$.
For two channels, $F_{\rm root}$ denotes the minimum unsquared fidelity
of their outputs on a purified input, retaining the reference system.
Then
\begin{equation}
 D_*=1-\max_{\omega_E}
 F_{\rm root}(\widehat{\cN},\mathcal C_{\omega_E})^2,
 \qquad \mathcal C_{\omega_E}(X)=\Tr(X)\omega_E,
 \label{eq:environmental-optimum}
\end{equation}
where $\omega_E$ is a density matrix on the environment, independent of
the input. B\'eny and Oreshkov's complementary-channel duality,
Corollary 5 of Ref.~\cite{BenyOreshkov2011}, identifies the maximized
squared fidelity with the optimal recovery fidelity. The Letter's
recovery equality gives Eq.~\eqref{eq:environmental-optimum}.
Thus the best fidelity approximation of the environment by a channel
that forgets the input determines the optimal operational recovery error.

\subsection{Approximate recovery}
Choose any state $\sigma$ on the logical input. Define the constant channel
$\mathcal S_\sigma(X)=\Tr(X)\sigma$ and let $c\in[0,1]$ be the
worst-input root fidelity between $\widehat{\cN}$ and the constant
environment channel $\widehat{\cN}\mathcal S_\sigma$.
B\'eny and Oreshkov use the distance
$d=\sqrt{1-F_{\rm root}}$. Since $\mathcal S_\sigma$ is an idempotent
generalized complementary channel of the identity, their
Corollary 6 and recovery construction,
Eq.~(75) of Ref.~\cite{BenyOreshkov2011}, give
\begin{equation}
 F(\cR_{\rm BO}\cN)\ge c^2,\qquad
 \beta:=\max_{\cR\ {\rm CPTP}}F(\cR\cN)
 \le\left(\frac{3+c}{4}\right)^2.
 \label{eq:bo-fidelity}
\end{equation}
Indeed, their optimal distance is at least $\sqrt{1-c}/2$, so
$1-\sqrt\beta\ge(1-c)/4$. The recovery $\cR_{\rm BO}$ is obtained
from their saddle-point construction.

Apply the Letter's unitary correction to $\cR_{\rm BO}\cN$ and use
$D_*=1-\beta$. Then
\begin{equation}
 1-\left(\frac{3+c}{4}\right)^2
 \le D_*\le D(\Ad_U\cR_{\rm BO}\cN)\le1-c^2\le4D_*.
 \label{eq:bo-operational}
\end{equation}
The lower bound on $D_*$ gives
$4D_*-(1-c^2)\ge3(1-c)^2/4\ge0$, proving the last inequality.
Thus one final unitary turns this established approximate recovery into
a dimension-independent factor-four approximation in diamond error.

\subsection{Accuracy of unitary-only correction}
\label{sec:sm-unitary-overhead}
Let $\cN$ be a channel on a $d$-dimensional system without a retained
noise record, with Kraus operators $A_j$. Write $e=\min_{\cR}D(\cR\cN)$
over all CPTP output corrections and $u=\min_U D(\Ad_U\cN)$ over unitaries.
We prove $u\le e/(1-e)$ and show that the leading coefficient in
$u-e\le e^2/(1-e)$ is sharp. The bound is automatic for $e\ge1/2$ since
$u\le1$. In finite dimension $e<1$, since the channel that prepares $I/d$
has positive worst-input entanglement fidelity.

Choose a fidelity-optimal correction $\cR$, with Kraus operators $R_i$.
The Letter gives $q=F(\cR\cN)=1-e>1/2$. At a minimizing state $\rho$,
the normalized coefficient matrix with entries
$\overline{\Tr(\rho R_i A_j)}/\sqrt q$ has Frobenius norm one.
The Letter's worst-input amplitude inequality gives
$\smin(T)^2\ge q$ for the combination $T$ with these coefficients.
By singular-value decomposition, rotate the two Kraus lists independently
and pad with zeros so that
\[
 T=\sum_i\sqrt{\lambda_i}R_i A_i,\qquad
 \lambda_1\ge\lambda_2\ge\cdots\ge0,\qquad\sum_i\lambda_i=1.
\]
The column of $A_i$ is an isometry. Multiplying it by the row of
$\sqrt{\lambda_i}R_i$ gives
$\sum_i\lambda_i R_iR_i^\dagger\succeq TT^\dagger\succeq qI$.
Taking traces and using $\sum_iR_i^\dagger R_i=I$ gives
$\lambda_1\ge q$. Put $\delta=1-\lambda_1$ and $c=\smin(R_1)^2$.
Trace preservation gives $R_i^\dagger R_i\preceq(1-c)I$ for $i>1$,
so the same bound holds for $R_iR_i^\dagger$. On a least left singular
vector of $R_1$ the row bound gives
\[
 q\le(1-\delta)c+\delta(1-c),\qquad
 c\ge\frac{q-\delta}{1-2\delta}\ge q.
\]
With $x=\smin(A_1)^2$, the squared norm of the row omitting its first
entry is at most $\delta(1-c)$, and that of the corresponding column
is at most $1-x$. Testing $T$ on a least right singular vector of
$A_1$ and using $\norm{R_1}\le1$ gives
\[
 \sqrt q\le\sqrt{(1-\delta)x}+\sqrt{\delta(1-c)(1-x)}
 \le\sqrt{x+(1-c)(1-x)}.
\]
Thus $1-x\le(1-q)/c\le(1-q)/q$. The inverse polar unitary of $A_1$
makes that Kraus operator positive with smallest eigenvalue $\sqrt x$.
Centering it by $\sqrt x I$ gives a centered Gram bounded by $(1-x)I$.
The Letter's diamond bound for this Gram gives $u\le1-x\le e/(1-e)$.

For sharpness, take qubit amplitude damping with probability
$0<\gamma<1/2$. The damped excited state has largest eigenvalue
$1-\gamma$, so testing for the excited state gives error at least
$\gamma$ after any unitary. The positive Kraus operator
$\diag(1,\sqrt{1-\gamma})$ attains this bound without correction, hence
$u=\gamma$. Now apply amplitude raising with probability
$p=\gamma-2\gamma^2$. Put $t=\sqrt{1-\gamma}$ and $v=\sqrt{1-p}$.
For an input marginal with excited population $y$, the squared
entanglement fidelity on the diagonal is
\[
 f(y)=[v+(t-v)y]^2+\gamma p y^2 .
\]
Off-diagonal entries of the marginal contribute only nonnegative
squared terms. The convex quadratic $f$ is decreasing on $[0,1]$
because $f'(1)/2=1-\gamma v^2-tv\le0$, as follows from
$(tv)^2-(1-\gamma v^2)^2=\gamma^3(4-5\gamma+4\gamma^2-4\gamma^3)>0$.
Its minimum is $f(1)=1-\gamma+\gamma^2-2\gamma^3$.
The fidelity--diamond equality and the upper bound on $u$ now give
$\gamma/(1+\gamma)\le e\le\gamma-\gamma^2+2\gamma^3$.
Consequently $(u-e)/e^2\to1$ as $\gamma\to0$.

\subsection{Circuit composition}
Let $\mathcal M_i$ be the CPTP map of gate $i$ and $\mathcal G_i$ its
ideal unitary channel. Put $F_i=F(\mathcal M_i\mathcal G_i^{-1})$.
Applying the Letter's correction to each relative channel gives
$\widetilde{\mathcal M}_i=\Ad_{U_i}\mathcal M_i$ with
$\norm{\widetilde{\mathcal M}_i-\mathcal G_i}_\diamond/2\le1-F_i$.
Consequently
\begin{equation}
 \frac12\norm{\widetilde{\mathcal M}_m\cdots\widetilde{\mathcal M}_1
                 -\mathcal G_m\cdots\mathcal G_1}_\diamond
 \le\min\left\{1,\sum_{i=1}^m(1-F_i)\right\}.
 \label{eq:circuit-budget}
\end{equation}
To prove this, telescope the difference of the two compositions,
replacing one gate at a time. CPTP preprocessing and postprocessing
cannot increase the diamond norm~\cite{Watrous2018}.
The bound therefore controls every final measurement probability,
including with an entangled reference.

\subsection{A fidelity-optimal decoder can require correction}

A short two-qubit counterexample establishes the strict distinction
between the sets of optimizing decoders. For $0<p<1/4$, put
$a=\sqrt{1-p}$ and $\theta=\arcsin\sqrt p$, and take
\begin{equation}
 A_0=\diag(e^{i\theta},e^{i\theta},a,a),\qquad
 A_1=\sqrt p\,\diag(0,0,1,-1).
\end{equation}
For $x=\rho_{00,00}+\rho_{01,01}$ and
$z=\rho_{10,10}-\rho_{11,11}$, the fidelity score is
$a^2+p(x^2+z^2)\ge1-p$, with equality at
$\rho_0=\proj1_A\otimes I_B/2$.
For any CPTP recovery, let $R_l$ be its blocks from the input $A=1$
sector to the output $A=1$ sector. They obey
$\sum_lR_l^\dagger R_l\preceq I_2$. Its fidelity on $\rho_0$ is
\begin{equation}
 \begin{split}
 &\frac14\sum_l[(1-p)|\Tr R_l|^2+p|\Tr(R_lZ)|^2]\\
 &\quad\le\frac{1-p}{2}\sum_l(|(R_l)_{00}|^2+|(R_l)_{11}|^2)
 \le1-p.
 \end{split}
\end{equation}
The first inequality uses $p\le\tfrac12$. Thus doing nothing already
maximizes worst-input fidelity over all CPTP recoveries.

Nevertheless, the product input $\ket+_A\ket0_B$ has output trace
distance $\sqrt p/2$ from its ideal value: its control coherence is
multiplied by $a e^{i\theta}$, with $|a e^{i\theta}-1|=\sqrt p$.
Hence $D(\Phi)\ge\sqrt p/2>p$. The weighted amplitude from $\rho_0$
is $T=aA_0$, whose polar correction
$U=\diag(e^{-i\theta},e^{-i\theta},1,1)$ makes the no-flag Kraus
operator $\diag(1,1,a,a)$ positive. The positive-Kraus bound gives
$D(\Ad_U\Phi)\le p$; the all-CPTP fidelity converse gives equality.
Both channels have optimal fidelity $1-p$, but only the corrected
channel is diamond optimal.

\section{Exact diamond error from coherent amplitudes}
\label{sec:sm-coherent}
Let $\Psi(X)=\sum_{j=1}^r C_jXC_j^\dagger$ and $D=D(\Psi)$.
The error has the exact representations
\begin{equation}
 \sqrt D=\inf_{\beta,\,z\ge\sqrt{1+\norm\beta_2^2}}
       \norm{zI-\sum_j\beta_jC_j},\qquad
 D=\inf_{\norm\gamma_2<1}
       \frac{\norm{I-\sum_j\gamma_jC_j}^2}{1-\norm\gamma_2^2}.
 \label{eq:sm-coherent}
\end{equation}
Here $\beta,\gamma\in\mathbb C^r$ and $z$ is real. Both infima are
attained for $D>0$.

For a purified input with marginal $\rho$, the output minus the ideal
pure input has at most one negative eigenvalue. Its magnitude $d(\rho)$
is the normalized trace distance. Put
$G_{ij}=\Tr(\rho C_i^\dagger C_j)$ and
$v_i=\Tr(\rho C_i^\dagger)$. The resolvent identity for this rank-one
perturbation, or its Schur complement, gives, for $a>0$,
\begin{equation}
 a\le d(\rho)\quad\Longleftrightarrow\quad
 M(\rho,a):=\begin{pmatrix}G+aI&v\\v^\dagger&1-a\end{pmatrix}\succeq0.
 \label{eq:sm-gram}
\end{equation}
At $a=0$, this is a Gram matrix and remains positive.
Maximizing $a$ over density matrices $\rho$ therefore gives $D$.
With $(K_1,\ldots,K_{r+1})=(C_1,\ldots,C_r,I)$, the dual is
\begin{equation}
 \min_{\substack{Z\succeq0\\Z_{r+1,r+1}-\Tr Z_{1:r,1:r}=1}}
 \lmax\!\left(\sum_{ij}Z_{ji}K_i^\dagger K_j\right).
 \label{eq:sm-gram-dual}
\end{equation}
For $D>0$, a full-rank input with positive error and
$0<a<d(\rho)$ gives strict primal feasibility. The dual is also strictly
feasible, so strong duality and attainment hold.
At any optimal input, $G+DI$ is invertible and $M(\rho,D)$ has a
one-dimensional kernel. Complementarity forces every optimal $Z$ to
have rank one. Write $Z=yy^\dagger$, with $y=(-\beta,z)$ and $z>0$.
The normalization becomes $z^2-\norm\beta_2^2=1$, and the objective is
$\norm{zI-\sum_j\beta_jC_j}^2$.
Scaling $(\beta,z)$ to the boundary cannot increase the objective;
setting $\gamma=\beta/z$ proves Eq.~\eqref{eq:sm-coherent} for $D>0$.
If $D=0$, all Kraus matrices are scalar multiples of $I$; aligned
coefficients tending to the cone boundary give the stated infimum.
No full-rank assumption on an optimal input is required.

\section{Complete optimal families and sharp gate radii}
\label{sec:sm-families}
Let $\Phi(X)=\sum_j A_jXA_j^\dagger$ be a fixed decoded channel with
$q=\max_U F(\Ad_U\Phi)$,
$0<q<1$, and put $t=\sqrt q$, $\epsilon=1-q$.
For a unit coefficient vector $c$, define $B_c=\sum_j\bar c_jA_j$.
For a fidelity-optimal gate $U$, stationarity at a worst marginal
gives $\operatorname{Re}(UB_c)\succeq tI$, where
$c_j=\Tr(\rho_0UA_j)/t$ and $\norm c_2=1$.
The minimizing amplitude vector is unique by the midpoint argument
in End Matter. Moreover $\smin(B_c)\ge t$; optimality of $q$ forces
equality. Conversely, $\operatorname{Re}(UB_c)\succeq tI$ implies
$\sum_j|\Tr(\rho UA_j)|^2\ge|\Tr(\rho UB_c)|^2\ge q$
for every $\rho$, hence fidelity optimality. Thus all such gates are
obtained by ranging over optimal coefficient rays.

Write $B_c=PS$, $tI\preceq S\preceq I$, and $U=WP^\dagger$.
For an operational optimizer, apply Eq.~\eqref{eq:sm-coherent} with
$\gamma=r\bar c$ and $b=\smin(B_c)$. The reverse triangle inequality
on a least-singular vector gives
\begin{equation}
 D\ge\frac{(1-rb)^2}{1-r^2}
   =1-b^2+\frac{(r-b)^2}{1-r^2}\ge1-q.
\end{equation}
Equality forces $r=b=t$. Expanding the residual norm then yields the
exact family constraints
\begin{equation}
 \operatorname{Re}(WS)\succeq T_F:=tI,\qquad
 \operatorname{Re}(WS)\succeq T_D:=\tfrac t2[(2-q)I+S^2],
 \label{eq:sm-families}
\end{equation}
for fidelity and operational optimality, respectively. The coefficient
in the second test is the gate's canonical coefficient: the stronger
inequality implies the first, and equality at a worst marginal makes
the amplitude vector $tc$. Phases must be chosen consistently:
replacing $U$ by $e^{i\phi}U$ replaces $c$ by $e^{i\phi}c$ and leaves
$UB_c$ unchanged.

Neither family contains a $W$ with eigenvalue $-1$. The Cayley
coordinate $W=(I+iH)(I-iH)^{-1}$, $H=H^\dagger$, changes either
condition, with $T=T_F$ or $T_D$, into
\begin{equation}
 S-T+i(HT-TH)-H(S+T)H\succeq0
 \quad\Longleftrightarrow\quad
 \begin{pmatrix}S-T+i(HT-TH)&H\\H&(S+T)^{-1}\end{pmatrix}\succeq0.
 \label{eq:sm-cayley}
\end{equation}
This is a convex matrix inequality in $H$.
Both families fix the space $E=\ker(S-tI)$ pointwise. On $E^\perp$,
the operational inequality has a strict margin at $H=0$, since
$a-T_D(a)=(a-t)[1-t(a+t)/2]>0$ for $t<a\le1$.
It therefore has nonzero interior whenever $E^\perp$ is nontrivial.

If $S=tI$, both families consist of the polar gate alone. Otherwise
choose an eigenvector of $S$ with eigenvalue $a>t$ and rotate it by
$\theta$, leaving its orthogonal complement fixed, with
$\cos\theta=t/a$. This gate is fidelity optimal, but
$|1-tae^{i\theta}|^2=\epsilon^2+q(a^2-q)>\epsilon^2$,
so it is not operationally optimal.
When $q>\tfrac12$, every optimal SDP density has rank one. Convexity then
forces a unique optimal density, since a mixture of two distinct pure
optima would have rank two. There is therefore just one coefficient
ray, proving the main text's global coincidence and uniqueness criterion.

To obtain physical gate radii, note that $T_F\succeq tS$ and
$T_D\succeq bS$, where $b=t(3-q)/2$, because
\begin{equation}
 T_D-bS=\tfrac t2(I-S)((2-q)I-S)\succeq0.
\end{equation}
Testing on an eigenvector of $W$ bounds its eigenphase by
$\arccos t$ or $\arccos b$. The unitary-channel distance is the sine
of half the smallest containing spectral arc when that arc is shorter
than $\pi$~\cite{Watrous2018}. Hence the radii are $\sqrt\epsilon$
and $\sqrt{1-b^2}=\epsilon\sqrt{3+\epsilon}/2$.
Both are sharp for amplitude damping on one qubit with a quiet second
qubit: take $A_0=\diag(1,1,t,t)$ and
$A_1=\sqrt\epsilon\ket0\bra1\otimes I_2$.
Testing $B_c^\dagger$ on $\ket{10}$ proves that the unique optimal
ray is $(1,0)$. The gates
$U_\theta=\diag(e^{i\theta},e^{-i\theta},1,1)$ saturate the two
radii at $\theta=\arccos t$ and $\theta=\arccos b$, respectively.

\section{Quadratic robustness of the polar correction}
\label{sec:sm-polar-robustness}
Let $\Lambda$ be the channel after an optimal polar correction.
Write $q=F(\Lambda)\in(0,1)$, $\epsilon=1-q$ and $t=\sqrt q$.
It has a Kraus operator $K$ with $tI\preceq K\preceq I$.
For a residual unitary $V$, let $s=d(V,I)<1$ and choose the scalar
phase of $V$ so that $\operatorname{Re}V\succeq cI$,
$c=\sqrt{1-s^2}$. For a unit vector $x$, set
$k=\langle x,Kx\rangle$ and $a=\norm{Kx}^2$.
Spectral calculus gives $t\le k\le1$ and
$k^2\le a\le(1+t)k-t$. Decomposing $Kx$ and $V^\dagger x$
into components parallel and orthogonal to $x$ gives
\begin{equation}
 \operatorname{Re}\langle x,VKx\rangle\ge ck-s\sqrt{a-k^2}.
\end{equation}
Thus $\norm{(I-rVK)x}^2\le1+r^2a-2rck+2rs\sqrt{a-k^2}$.
For $r\ge0$ this increases with $a$. Substituting its upper bound
and writing $k=t+(1-t)u$ identifies a two-dimensional Rayleigh quotient:
\begin{equation}
 M_{t,s}(r)=\begin{pmatrix}
 1+r^2t^2-2rct&rs(1-t)\\rs(1-t)&1+r^2-2rc
 \end{pmatrix},\qquad
 D(\Ad_V\Lambda)\le\mathfrak B_q(s):=
 \inf_{0\le r<1}\frac{\lmax M_{t,s}(r)}{1-r^2}.
 \label{eq:sm-polar-envelope}
\end{equation}
The last step uses Eq.~\eqref{eq:sm-coherent} in the $K$ direction.
Each supplied $r$ gives a finite certificate. For $s=1$, $r=0$
gives the bound one.

The entire envelope is attained on three logical states by
\begin{equation}
 K=\diag(t,1,1),\quad L=\sqrt\epsilon\ket2\bra0,\qquad
 V=\begin{pmatrix}c&-s&0\\s&c&0\\0&0&1\end{pmatrix}.
 \label{eq:sm-polar-sharp}
\end{equation}
Testing $B_c^\dagger$ on $\ket0$ bounds its least singular value by
$t$, so the nominal optimal fidelity is exactly $q$.
In Eq.~\eqref{eq:sm-coherent}, pinching onto $\{0,1\}$ and $\{2\}$
removes the coefficient of $L$, decreasing the numerator and enlarging
the denominator. Conjugation symmetry and convexity of the level test
$\norm{I-\gamma VK}^2+D|\gamma|^2\le D$ allow a real coefficient.
Negative coefficients give a quotient at least one. The spectator
residual $|1-r|$ never exceeds the two-dimensional block norm, whose
square is $\lmax M_{t,s}(r)$. This proves exact equality with the envelope.

At $s=0$, the unique minimum is $r=t$. The top eigenvalue is
$\epsilon^2$, with gap $t(t+2)(1-t)^2>0$.
Analytic eigenvalue perturbation and the nondegenerate scalar minimum
give $r=t+O_q(s^2)$. At $r=t$ the second-order eigenvalue coefficient
is $t^2+t/(t+2)=t(1+t)^2/(t+2)$. Therefore
\begin{equation}
 \mathfrak B_q(s)=\epsilon+
 \frac{t(1+t)^2}{(t+2)\epsilon}s^2+O_q(s^4).
 \label{eq:sm-polar-expansion}
\end{equation}
The expansion fixes $0<q<1$; at $q=1$ the envelope is $s$.

\section{Sharp four-thirds robustness of operationally optimal gates}
\label{sec:sm-critical}
Suppose $\Lambda$ already has optimal final-unitary error
$D(\Lambda)=\epsilon=1-q$, with $0<q<1$, and put $t=\sqrt q$.
Its optimal coherent amplitude $B$ obeys
$B^\dagger B\succeq qI$, $\norm B\le1$, and the operational inequality
in Eq.~\eqref{eq:sm-families}. Set $S_0=B^\dagger B-qI$ and
$C=B-tI$. Expansion gives $C^\dagger C\preceq\epsilon S_0$.
For $r=t-\eta$, $0<\eta\le t$, the radial residual has margin
\begin{equation}
 (1-rt)^2I-(I-rB)^\dagger(I-rB)
 =2r\operatorname{Re}(B-tI)-r^2S_0\succeq r\eta S_0.
\end{equation}
For a residual unitary $V$, put $\Delta=V-I$ and $e=\norm\Delta$, choosing the irrelevant phase
of $V$ to minimize $e$. The disturbed residual Gram changes by
$-2r\operatorname{Re}(\Delta B)$.
Its scalar part is at most $rt e^2I$, since
$\Delta+\Delta^\dagger=-\Delta^\dagger\Delta$.
For the other part, Young's inequality gives
\begin{equation}
 -2r\operatorname{Re}\langle x,\Delta Cx\rangle
 \le2re\sqrt\epsilon\sqrt{\langle x,S_0x\rangle}
 \le r\eta\langle x,S_0x\rangle+r\epsilon e^2/\eta.
\end{equation}
The first term consumes the radial margin. Dividing the remaining
bound by $1-r^2$ in Eq.~\eqref{eq:sm-coherent} yields the finite bound
in End Matter. In particular,
\begin{equation}
 0\le D(\Ad_V\Lambda)-\epsilon
 \le\min\left\{q,\,
 3\left(\frac q{4\epsilon}\right)^{1/3}e^{4/3}
             +\frac q\epsilon e^2\right\}.
 \label{eq:sm-critical-bound}
\end{equation}
Indeed, $1-r^2\ge\epsilon$ bounds the excess by
$\eta^2/\epsilon+te^2/\eta+qe^2/\epsilon$.
Choose $\eta=(t\epsilon e^2/2)^{1/3}$ when this is at most $t$;
otherwise the displayed expression already exceeds the trivial bound $q$.
For $s=d(V,I)<1$, $e=\sqrt{2-2\sqrt{1-s^2}}=s+O(s^3)$.
Output unitary drift leaves the optimal final-unitary value unchanged,
so this excess is also the regret of retaining the nominal gate.

\emph{Sharpness.}
For any fixed $0<q<1$, put
\begin{equation}
 \begin{gathered}
 u=\tfrac t2(3-q),\quad v=\tfrac\epsilon2\sqrt{4-q},\quad z=u+iv,\\
 K=\diag(t,z,1),\quad L=\sqrt\epsilon\ket2\bra0,\qquad
 V_s=\begin{pmatrix}\sqrt{1-s^2}&-is&0\\-is&\sqrt{1-s^2}&0\\0&0&1\end{pmatrix}.
 \end{gathered}
 \label{eq:sm-critical-channel}
\end{equation}
Here $|z|=1$. The nominal channel $\Phi$ is trace preserving and has
optimal fidelity $q$, by the same least-singular-value test as above.
Both the identity and $P=\diag(1,\bar z,1)$ attain error $\epsilon$;
the identity saturates the operational constraint on its second level.
For the unchanged identity correction we will prove
\begin{equation}
 D(\Ad_{V_s}\Phi)-\epsilon
   =3\left(\frac{q^2}{16\epsilon}\right)^{1/3}s^{4/3}
      +o_q(s^{4/3}).
 \label{eq:sm-critical-asymptotic}
\end{equation}
For the unchanged polar correction on the same physical channel,
$D(\Ad_{PV_s}\Phi)-\epsilon$ instead equals the quadratic expansion
in Eq.~\eqref{eq:sm-polar-expansion}: move $P$ through $V_s$ by
conjugation, then remove the two-level rotation phase by a diagonal
basis change that preserves the positive $K$ and the jump channel.

To prove Eq.~\eqref{eq:sm-critical-asymptotic}, pinching again removes
the jump coefficient. Write $c=\sqrt{1-s^2}$ and let
$G=(I-\gamma V_sK)^\dagger(I-\gamma V_sK)$ on the first two levels.
Its exact entries are
\begin{equation}
 \begin{split}
 G_{00}&=1+q|\gamma|^2-2tc\operatorname{Re}\gamma,\\
 G_{11}&=1+|\gamma|^2-2c\operatorname{Re}(\gamma z),\\
 G_{01}&=is(\gamma z-t\bar\gamma).
 \end{split}
 \label{eq:sm-critical-gram}
\end{equation}
Put $D=\epsilon+h$ and $\gamma=t+x+iy$. The diagonal entries of the
slack $(\epsilon+h)(1-|\gamma|^2)I-G$ are
\begin{equation}
 \begin{split}
 A&=h(1-|\gamma|^2)-x^2-y^2-2t(1-c)(t+x),\\
 C_1&=-t\epsilon x-2vy+h(1-|\gamma|^2)
       -(2-q)(x^2+y^2)+2(c-1)\operatorname{Re}(\gamma z).
 \end{split}
 \label{eq:sm-critical-slack}
\end{equation}
The universal bound gives $h=O_q(s^{4/3})$. At an optimum below one,
$G_{00}\le D(1-|\gamma|^2)$ forces $\operatorname{Re}\gamma>0$.
Then $A\ge0$ gives $|\gamma-t|^2\le h$, so $x,y=O_q(s^{2/3})$.
The spectator $|1-\gamma|$ is strictly inactive in this limit, since
$1-t<\epsilon$. Moreover
$|\gamma z-t\bar\gamma|^2\to q\epsilon^2>0$.
The determinant constraint $AC_1\ge s^2|\gamma z-t\bar\gamma|^2$
therefore forces $C_1$ to be of order at least $s^{2/3}$.

Set $\delta=s^{2/3}$, $w=(x+iy)/\delta$, $H=h/\delta^2$, and
$\ell=(-t\epsilon,-2v)$, so $\norm\ell_2=2\epsilon$.
The scaled optimizers are bounded. Every subsequential feasible limit
obeys
\begin{equation}
 (\epsilon H-|w|^2)\,\ell\cdot(\operatorname{Re}w,\operatorname{Im}w)
       \ge q\epsilon^2,
 \qquad \epsilon H-|w|^2>0,
\end{equation}
with positive dot product on the left.
For $a=|w|>0$, Cauchy--Schwarz implies
\begin{equation}
 H\ge\frac{a^2}{\epsilon}+\frac q{2a}
   \ge\frac{3R^2}{\epsilon},\qquad R=(q\epsilon/4)^{1/3}.
\end{equation}
The last minimum follows from
$a^3+2R^3-3aR^2=(a-R)^2(a+2R)\ge0$.
For a matching upper limit take
$w_*=R(-t/2-i\sqrt{4-q}/2)$, $\gamma=t+\delta w_*$ and
$H=3R^2/\epsilon+\zeta$, with fixed $\zeta>0$.
The leading diagonal slacks are positive and the leading determinant
has positive margin $2\epsilon^2R\zeta$. Thus the exact slack is
positive for all sufficiently small $s$. Letting $\zeta$ decrease to
zero proves the matching limit and Eq.~\eqref{eq:sm-critical-asymptotic}.
The bound in Eq.~\eqref{eq:sm-critical-bound} is therefore sharp
in its exponent $4/3$.

\section{Finite-precision certificates and implementation error}
\label{sec:sm-precision}

For a nominal decoded channel $\Phi_0(X)=\sum_jA_jXA_j^\dagger$, choose
any density matrix $\rho$. Put
\begin{equation}
 \alpha_j=\Tr(\rho A_j),\quad a=\sum_j|\alpha_j|^2,\quad
 T=\sum_j\bar\alpha_j A_j,\quad m=\lmin(\operatorname{Re}T).
\end{equation}
Convexity supplies a two-sided check on the minimum:
\begin{equation}
 \max(0,2m-a)\le F(\Phi_0)\le a.
\end{equation}
Indeed, the quadratic score at $\sigma$ is at least its tangent value
$2\Tr(\sigma\operatorname{Re}T)-a$. If $a>0$ and $UT=|T|$, a Kraus
rotation selects $K=|T|/\sqrt a$. The positive-Kraus bound in End Matter gives
\begin{equation}
 D(\Ad_U\Phi_0)\le1-\frac{\smin(T)^2}{a}.
 \label{eq:residual}
\end{equation}
With validated bounds
$a\le\bar a$ and $\smin(T)\ge\underline s\ge0$, the upper bound
$1-\underline s^2/\bar a$ remains valid for $\bar a>0$.
At $a=0$ use $D\le1$. Exact arithmetic or spectral error enclosures
turn these bounds into rigorous finite-error certificates.

An exactly specified unitary $Q$ can be certified without estimating its
distance to a potentially unstable polar factor. For any coefficient
vector $\gamma$ with $\norm\gamma_2<1$, Eq.~\eqref{eq:sm-coherent} gives
\begin{equation}
 D(\Ad_Q\Phi_0)\le
 c_Q:=\min\left\{1,
 \frac{\norm{I-\sum_j\gamma_j QA_j}^2}{1-\norm\gamma_2^2}\right\}.
\end{equation}
Each supplied $\gamma$ gives an upper bound; its infimum over $\gamma$
is the exact nominal error.

Suppose the actual logical channel $\Phi$ and implemented correction
$\mathcal K$ satisfy normalized radii
$\norm{\Phi-\Phi_0}_\diamond/2\le\delta$ and
$\norm{\mathcal K-\Ad_Q}_\diamond/2\le\epsilon_{\rm gate}$. Contractivity gives
\begin{equation}
 \begin{split}
 D(\mathcal K\Phi)
 &\le\min\{1,D(\Ad_Q\Phi_0)+\delta+\epsilon_{\rm gate}\}\\
 &\le\min(1,c_Q+\delta+\epsilon_{\rm gate}).
 \end{split}
 \label{eq:implementation-budget}
\end{equation}
For a polar correction, Eq.~\eqref{eq:residual} supplies an alternative
nominal bound.
A radius $\delta$ for the physical noise also bounds the logical channel
after composition with a fixed CPTP decoder.

For a finite final-unitary optimality certificate, supply a normalized
coherent direction $c$ and an input state $\sigma$ such that
$\widehat\Phi_0(\sigma)\preceq uI$.
Write $D_U^*=\min_U D(\Ad_U\Phi_0)=1-q_*$ for the optimal final-unitary
error. With $q=\smin(B_c)^2$,
\begin{equation}
 q\le q_*\le\Gamma\le u,\quad
 1-u\le D_U^*\le1-q,\quad
 D(\Ad_{U_c}\Phi_0)-D_U^*\le u-q.
\end{equation}
This bracket is valid at every error, even when the first SDP is not
exact. For the actual channel, the nominal optimum changes by at most
$\delta$, so the implemented correction's regret is at most
$u-q+2\delta+\epsilon_{\rm gate}$.

\section{All-error certificates from symmetric extensions}
\label{sec:sm-hierarchy}

The Letter's SDP is exact whenever the best final-unitary error is below
$\tfrac12$. At larger errors
the coherent-direction problem is rank constrained. The hierarchy below
specializes the established bosonic framework of Yu \emph{et al.}~\cite{Yu2022}.
Let $E=\mathbb C^r$, let $L=\widehat\Phi^\dagger$, and at level $k$
choose a density matrix $\Omega_k$ supported on $\operatorname{Sym}^k(E)$.
Write $\Omega_{k-1}=\Tr_1\Omega_k$ and define
\begin{equation}
 \Gamma_k=\max\{t\in[0,1]:
 (L\otimes\id)(\Omega_k)\succeq tI_d\otimes\Omega_{k-1}\}.
\end{equation}
At fixed $t$ the constraint is affine in $\Omega_k$, so monotone
bisection computes $\Gamma_k$ through SDP feasibility problems.
Product states and partial tracing give
$\Gamma_1\ge\Gamma_2\ge\cdots\ge q_*$, with $\Gamma_1=\Gamma$.

Put $m=r-1$. If $\Gamma_1>\tfrac12$ all levels are already exact.
Otherwise, for $k>m$,
\begin{equation}
 0\le\Gamma_k-q_*\le\frac{m(1-2q_*)}{k+m},\qquad
 \lim_k\Gamma_k=q_*.
 \label{eq:hierarchy-rate}
\end{equation}
Here is a short constructive argument. Project the last $k-1$ factors
onto $\psi^{\otimes(k-1)}$, leaving $X_\psi\succeq0$, with
$x_\psi=\Tr X_\psi$. Feasibility implies
$L(X_\psi)\succeq t x_\psi I$. If
$D_j=\dim\operatorname{Sym}^j(E)$, Haar moments give
$\int x_\psi\,d\psi=1/D_{k-1}$ and
$\int\bra\psi X_\psi\ket\psi\,d\psi=1/D_k$.
Thus some normalized conditional state has leading eigenvalue
$\lambda\ge D_{k-1}/D_k=k/(k+m)>\tfrac12$.
The same spectral rounding as in End Matter gives
$t\le1-\lambda+(2\lambda-1)q(c)$ for its leading eigenvector $c$.
For $t\le\tfrac12$ this yields
\begin{equation}
 q(c)\ge\frac{(k+m)t-m}{k-m}.
\end{equation}
Using $q(c)\le q_*$ at $t=\Gamma_k$ gives
Eq.~\eqref{eq:hierarchy-rate}. A finite positive projective $k$-design
reproduces the two moments and makes the extraction finite. Polar
correction then translates it into a diamond-error certificate.
The semidefinite variable has dimension $\binom{k+r-1}{r-1}$,
which grows with the level $k$ and Kraus rank $r$.

\section{Three-qubit conditional-gate calibration}
\label{sec:sm-calibration}

\subsection{Average-fidelity and leading-Kraus comparators}

For a $d$-level system, maximizing average gate fidelity is equivalent
to maximizing tracial entanglement fidelity~\cite{Nielsen2002}:
\begin{equation}
 f_{\rm tr}(\mathcal R\Phi)=\frac1{d^2}\sum_{l,j}|\Tr(R_lA_j)|^2,
 \qquad F_{\rm av}=\frac{d f_{\rm tr}+1}{d+1},
\end{equation}
where $R_l$ are recovery Kraus operators. This objective fixes the
logical marginal to $I/d$. Write $m=d-2$ for the number of quiet levels;
the three-qubit model has $d=8$, $m=6$. Its Kraus operators are
$A_0=\diag(I_m,a,a)$ and $A_1=\sqrt p\,\diag(0_m,1,i)$,
with $a=\sqrt{1-p}$.
The family $U_\alpha=\diag(I_m,e^{i\alpha},e^{-i\alpha})$ has objective
\begin{equation}
 f_{\rm tr}(\Ad_{U_\alpha}\Phi)
 =\frac{(m+2a\cos\alpha)^2+2p(\cos\alpha+\sin\alpha)^2}{d^2}.
\end{equation}
Its derivative vanishes at
\begin{equation}
 p\cos(2\alpha)=a(m+2a\cos\alpha)\sin\alpha.
 \label{eq:average-stationary}
\end{equation}
For $0<p<1$, on $[0,\pi/4]$ the left side strictly decreases and the
right side strictly increases, with opposite endpoint signs of their
difference. The solution $\alpha_{\rm av}$ is therefore unique.

To certify optimality over all CPTP recoveries, define
$C_{kl}=\sum_j\overline{(A_j)_{kk}}(A_j)_{ll}$ and let
$r_l$ collect the diagonal entries of $R_l$. Trace preservation gives
$\sum_l|(r_l)_k|^2\le1$, whereas
$d^2f_{\rm tr}=\sum_l r_l^\dagger C r_l$. A positive diagonal matrix
$Y\succeq C$ therefore supplies the upper bound $d^2f_{\rm tr}\le\Tr Y$.
At $\alpha=\alpha_{\rm av}$, set
\begin{equation}
 \begin{split}
 h&=m+2a\cos\alpha,\\
 k&=ma\cos\alpha+a^2\cos(2\alpha)+p\sin(2\alpha),\\
 Y&=\diag(hI_m,1+k,1+k).
 \end{split}
\end{equation}
The slack $Y-C$ annihilates
$u=(1_m,e^{i\alpha},e^{-i\alpha})$, where $1_m$ is the all-ones vector.
On the quiet subspace orthogonal
to $1_m$ it equals $hI$. In the remaining three-dimensional space,
the quiet diagonal entry is $2a\cos\alpha>0$. Its Schur complement is
\begin{equation}
 \beta\begin{pmatrix}1&-e^{2i\alpha}\\-e^{-2i\alpha}&1\end{pmatrix},
 \quad \beta=k-\frac{ma}{2\cos\alpha}.
\end{equation}
Equation~\eqref{eq:average-stationary} gives
$\beta e^{2i\alpha}=a^2+ma/(2\cos\alpha)+ip$, so
$\beta=(a^2+ma/(2\cos\alpha))\cos(2\alpha)+p\sin(2\alpha)>0$.
Thus $Y-C\succeq0$ with kernel $\operatorname{span}(u)$, and
$U_\alpha$ attains the dual bound. Equality forces each $r_l$ to be
proportional to $u$ and every diagonal trace-preservation bound to
saturate. All off-diagonal recovery entries then vanish, so
$R_l=\lambda_lU_\alpha$ and the recovery channel is unique.
The endpoint choices are $0,\pi/4$ at $p=0,1$. At weak noise,
$\alpha_{\rm av}\sim p/d$; the worst-case angle is $\alpha_*\sim p/2$.

For the prior leading-Kraus polar correction~\cite{Carignan2019}, diagonalize
$G_{ij}=\Tr(A_iA_j^\dagger)/d$, choose a normalized leading eigenvector
$c$, and undo the polar unitary of $B_c=\sum_j\bar c_jA_j$.
This uses the same input channel and final-unitary access as both optima.

\subsection{Repairing the average-calibrated channel}

For the three-qubit channel and a correction $U_\alpha$, the Kraus
traces depend only on diagonal input populations. A common phase change
of $A_1$ makes the two active amplitude vectors complex conjugates.
Exchanging their populations therefore conjugates both traces; convexity
shows that the minimum occurs with equal active populations.
If their total population is $x$, define
$b=1-a\cos\alpha$ and $k=p[1+\sin(2\alpha)]/2$. For $b^2+k>0$,
\begin{equation}
 F(\Ad_{U_\alpha}\Phi)=\min_{0\le x\le1}[(1-bx)^2+kx^2],
 \qquad x_{\min}=\min\{1,b/(b^2+k)\}.
 \label{eq:calibration-fidelity}
\end{equation}
If $b=k=0$, the channel is ideal and $F=1$.
At $p=.01$ and $\alpha=\alpha_{\rm av}$, this gives
$x_{\min}=.9951640112$ and $F=.99501089446$.
Use the marginal
\[
 \rho_0=\diag([(1-x_{\min})/6]I_6,x_{\min}/2,x_{\min}/2)
\]
in the main theorem's weighted polar construction, applied to the
already calibrated channel. The resulting total angle is
$.00500704151$, compared with $\alpha_*=.00500008334$.
Its diamond error has numerical lower and upper bounds
$.00497525771$ and $.00497525807$; the exact optimum is $.00497524814$.
The theorem itself certifies the bound $1-F=.00498910554$.

For the product input $\ket{11+}$ the target off-diagonal matrix element
is $z_\alpha/2$, where $z_\alpha=(1-p-ip)e^{2i\alpha}$.
Thus $\langle X\rangle=\operatorname{Re}z_\alpha$ and
$\langle Y\rangle=-\operatorname{Im}z_\alpha$. This gives the local
readout in the Letter.

\subsection{Compilation and implementation noise}

With $R_z(t)=e^{-itZ/2}$ and $\beta=-2\alpha$, each correction has
this time-ordered sequence on target $B$:
\begin{equation}
 \begin{gathered}
 R_z(\beta/4),\ \mathrm{CX}_{C_2,B},\ R_z(-\beta/4),\ \mathrm{CX}_{C_1,B},\\
 R_z(\beta/4),\ \mathrm{CX}_{C_2,B},\ R_z(-\beta/4),\ \mathrm{CX}_{C_1,B}.
 \end{gathered}
\end{equation}
For fixed control bits $c_1,c_2$, the net rotation angle is
$\beta[1-(-1)^{c_2}+(-1)^{c_1+c_2}-(-1)^{c_1}]/4$.
It is $\beta$ for $c_1=c_2=1$ and zero otherwise. This verifies a
four-CNOT, four-rotation implementation for all three comparators.

Append independent $Z$ flips of probability $\lambda$ to each data
qubit after the conditional noise. After every correction gate, apply
independent $Z$ flips of probability $\epsilon$ to its participating
qubits: one for each rotation and two for each CNOT, giving 12 locations.
CNOT propagation maps these errors to diagonal Pauli operators; rotations
commute with them. Hence the compiled correction equals $U_\alpha$
followed by a diagonal noise channel independent of $\alpha$.

For a robustness example, take $p=.01$ and $\lambda=\epsilon=10^{-4}$.
Reoptimize average gate fidelity within this same $U_\alpha$ circuit
family and compare with the fixed ideal-model angle $\alpha_*$.
To specify the reoptimization, let $G$ be the Schur symbol
of the channel including background and propagated implementation noise,
before applying $U_\alpha$. Up to a positive factor and constant, the
average objective is
\begin{equation}
 A\cos\alpha+B\cos(2\alpha)+C\sin(2\alpha),
\end{equation}
where $A=2\sum_{i=0}^5(G_{i6}+G_{i7})$, $B=2\operatorname{Re}G_{67}$,
and $C=-2\operatorname{Im}G_{67}$. These coefficients are real and
nonnegative in this noise model. The maximum is the unique stationary
point on $[0,\pi/4]$.

The normalized diamond errors, including background and correction-gate
noise, are
\begin{equation}
 \label{eq:implementation-example}
 \begin{aligned}
 D_{\rm no\ correction}&=.00714071,\\
 D_{\rm average\ calibration}&=.00697237,\\
 D_{\alpha_*}&=.00613394.
 \end{aligned}
\end{equation}
The uncorrected channel includes the conditional and background noise. The correction
reduces error by $12.0\%$ relative to reoptimized average calibration and
$14.1\%$ relative to no correction. For other implementation models,
Eq.~\eqref{eq:implementation-budget} supplies the corresponding additive
error budget.

For these diagonal channels, write $H=G-\bm1\bm1^T$ for the
Schur symbol of the difference from identity. Diagonal-unitary covariance
reduces the diamond-norm program to
\begin{equation}
 D=\frac12\max_{r,W}\{\Tr(HW):r\ge0,\ \bm1^Tr=1,
       -\diag(r)\preceq W\preceq\diag(r)\}.
\end{equation}
Twirling diagonalizes the input marginal. Compressing the primal matrix
to $\operatorname{span}\{\ket{ii}\}$ retains the objective, and zero
extension gives the converse feasibility implication. All eight data
levels remain, including background errors in the quiet sector.
Physical input trace norms and the dual constraints $Z\succeq0$,
$Z\succeq H$ with objective $\max_i Z_{ii}$ provide independent numerical
bounds. For the three errors in Eq.~\eqref{eq:implementation-example},
the gaps between these bounds are below $8\times10^{-10}$.

\subsection{Coherent-calibration tolerance}
\label{sec:sm-tolerance}
When the optimal coherent amplitude is scalar after correction,
a Kraus change gives $\Psi=q\,\id+\epsilon\mathcal N$,
where $q$ is the optimal fidelity, $\epsilon=1-q$ and $\mathcal N$ is CPTP.
This applies to the Letter's three-qubit
example by its exact binary-mixture certificate.
For a purified input $\ket\psi$ and residual unitary $W$, write
$\ket v=(W\otimes I)\ket\psi$. The actual output minus the ideal
input is $q\proj v-\proj\psi+\epsilon\sigma$, with $\sigma\succeq0$.
The first two terms have one negative eigenvalue, of magnitude
\begin{equation}
 \frac{\epsilon+\sqrt{\epsilon^2+4q(1-|\langle\psi|v\rangle|^2)}}2.
\end{equation}
Adding the positive matrix $\epsilon\sigma$ cannot lower that
eigenvalue. The full trace-zero difference has at most one negative
eigenvalue, so its normalized trace norm is bounded by this magnitude.
Since $1-|\langle\psi|v\rangle|^2\le s^2$ for $s=d(W,I)$,
\begin{equation}
 D(\Ad_W\Psi)\le\frac{\epsilon+\sqrt{\epsilon^2+4q s^2}}2,
 \qquad
 s\le\sqrt{\frac{\eta(\epsilon+\eta)}q}
 \ \Longrightarrow\ D(\Ad_W\Psi)\le\epsilon+\eta.
 \label{eq:sm-tolerance}
\end{equation}
For $p=0.01$, $\epsilon=0.0049752481\ldots$ and $\eta=0.1\epsilon$,
the allowed $s$ is $0.0016542\ldots$; the main text rounds down.
This is a guarantee for coherent gate error. An additional channel-model
or nonunitary implementation error adds by Eq.~\eqref{eq:implementation-budget}.

\section{Certification of the attainable region in Fig. 1(b)}
\label{sec:sm-region}

The region ranges over all output unitaries on the three data qubits.
At $p=0.01$, use the corrected frame from the Letter,
\[
 \begin{gathered}
 \Psi=q\,\mathrm{id}+e\,\Ad_V,\qquad
 q=\frac{1+\sqrt{99/101}}2,\quad e=1-q,\\
 V=\diag(-I_6,p-i\sqrt{1-p^2},p+i\sqrt{1-p^2}).
 \end{gathered}
\]
A fixed change of output frame bijects the unitary corrections and hence
preserves their performance region. With
$\varepsilon=1-F(\Ad_U\Psi)$, let $D_-(\varepsilon)$ and
$D_+(\varepsilon)$ be the minimum and maximum of $D(\Ad_U\Psi)$ at
that infidelity. The figure covers $e\le\varepsilon\le0.00501$.

\emph{Geometry between sampled fidelities.}
For $e\le\varepsilon<1/4$, every value between $D_-$ and $D_+$ is
attained, and both boundary functions are continuous and nondecreasing.
We summarize the geometric argument that permits certification of the
continuum from endpoint enclosures.

For Kraus operators $A_j$ of $\Psi$, write
$B_c=\sum_j\bar c_jA_j$. For a threshold $f>1/2$, the condition
$F(\Ad_U\Psi)\ge f$ is equivalent
to the existence of a unit coefficient $c$ with
$\operatorname{Re}(UB_c)\succeq\sqrt f I$.
Writing $B_c=PS$, the Cayley coordinate $H$ satisfies
\[
 H(S+\sqrt f I)H\preceq S-\sqrt f I.
\]
Scaling $H$ to zero contracts each fiber to its polar correction.
The feasible coefficient rays also contract: interpolate their rank-one
states to a fixed optimal coefficient state and take the leading
eigenvector. If its eigenvalue is $\lambda$, the spectral rounding
inequality from the Letter's SDP proof gives
$B_c^\dagger B_c\succeq(f+\lambda-1)/(2\lambda-1)\,I\succeq fI$.
Since $\lambda\ge f>1/2$, this operation is continuous.
The minimizing Kraus-trace vector is unique and continuous, so these
contractions define a contraction of the whole fidelity superlevel in
$PU(8)$.

Every point at fidelity $f>1/2$ admits a perturbation to lower fidelity.
Indeed, for a worst marginal $\rho$ and a Hilbert--Schmidt orthonormal
Hermitian basis $H_a$, the fixed-input tests
$g_a(s)=\sum_j|\Tr(\rho e^{isH_a}UA_j)|^2$ obey
\[
 \sum_a g_a''(0)=2(\Tr\rho^2-8f)<0.
\]
Thus the exact level is the boundary of the contractible superlevel.
These sets are semialgebraic. Duality in the closed orientable manifold
$PU(8)$ implies that the complement of the contractible superlevel is
connected; the Mayer--Vietoris sequence and $H^1(PU(8);\mathbb Z)=0$
then imply that its boundary is connected. Semialgebraic connectedness
gives path connectedness. Continuity of $D$ therefore fills each vertical
interval. Below the optimal fidelity, small moves of the coefficient
toward its optimum, combined with shrinking $H$, give nearby higher
fidelity. Together with the lower perturbation, this makes the compact
exact levels, and their extremal errors, continuous in $f$.

The same contraction using the optimal coherent coefficient in
Eq.~\eqref{eq:sm-coherent} connects each correction with $D<1/2$ to an
optimal polar correction without increasing $D$. Passing through
intermediate fidelity levels proves monotonicity of $D_-$.
For $D_+$, fix a maximizing input and its rank-one Helstrom projector
$M$, and write $\sigma$ for the noisy joint output and $\rho_R$ for its
reference marginal. The sum of second derivatives of the fixed
measurement test under output perturbations is
\[
 2\{8\Tr(M\sigma)-\Tr[M(I\otimes\rho_R)]\}
 \ge 8(1-D)-2>0
\]
when $D<3/4$. Here positivity of $\sigma$ gives
$\Tr(M\sigma)\ge(1-D)/2$ and
$\Tr[M(I\otimes\rho_R)]\le1$.
Thus $D$ has no interior local maximum in this range. Its maximum over
$F\ge f$ lies at $F=f$, and nesting proves monotonicity of $D_+$.
The universal bound $D\le\sqrt\varepsilon<1/2$ ensures that these
arguments apply throughout the stated range.

\emph{Global endpoint bounds.}
The coefficient search includes noncommuting corrections and arbitrary
worst-input ranks. In the corrected frame,
$B_c=\sqrt q\,\bar c_0I+\sqrt e\,\bar c_1V$.
Fixing the coefficient phase so that $c_0\ge0$, the faithful marginal
with $\Tr(\rho_*V)=0$ gives
\[
 |c_1|^2\le\frac{q-f}{q-e}.
\]
A complete adaptive rectangle cover of this coefficient disk supplies
the upper bound on $D_+$. On each rectangle, perturbation margins turn
$\operatorname{Re}(UB_c)\succeq\sqrt f I$ into sectorwise matrix
inequalities valid throughout the rectangle. Positive semidefinite dual
certificates for the resulting three-sector moment relaxation bound
$D$ for every admissible unitary. An independent checker verifies the
complete cover, exclusions and dual matrix positivity using outward
interval arithmetic, including every LDL pivot. Solver output supplies
candidate certificates; the interval checks establish their validity.

A lower bound on the other edge is
\[
 D_-(\varepsilon)\ge\sqrt{\varepsilon-qe}.
\]
For a purified input, the output has rank at most two, with nonzero
eigenvalue product at most $qe$. If the ideal-minus-noisy output has
eigenvalues $\delta,-a,-b$, then
$\delta^2-ab=1-\langle\psi|\sigma|\psi\rangle
 -\lambda_1(\sigma)\lambda_2(\sigma)$.
Applying this at a fidelity-worst input proves the bound.

Inner witnesses are explicit phase corrections of the original channel,
with three quiet phases $+\theta$, three $-\theta$, and active phases
$\pm\alpha$. Put $h=\cos\theta$, $a=\sqrt{1-p}\cos\alpha$ and
$k=p[1+\sin(2\alpha)]/2$. Their fidelity is exactly
\[
 F=\min_{0\le x\le1}\{[h-(h-a)x]^2+kx^2\}.
\]
Solving this expression for $h$ at each target $f$ gives witnesses at
exact fidelity. Equation~\eqref{eq:sm-coherent} bounds the lower witness's
diamond error from above; explicit inputs and rank-one measurements
bound the upper witness's error from below.

\emph{Enclosing the full plotted region.}
At 840 nodes $\varepsilon_i$, the checks give
$L_i\le D_-(\varepsilon_i)\le A_i$ and
$B_i\le D_+(\varepsilon_i)\le C_i$.
Monotonicity then gives, for
$\varepsilon_i\le\varepsilon\le\varepsilon_{i+1}$,
\[
 L_i\le D_-(\varepsilon)\le A_{i+1},\qquad
 B_i\le D_+(\varepsilon)\le C_{i+1}.
\]
The shaded rectangles lie between $A_{i+1}$ and $B_i$ and are attained.
Seventy-decimal outward interval checks give maximum enclosure widths
below $3.431\times10^{-5}$ and $4.77\times10^{-5}$, respectively, in
normalized diamond error. These certify both edges to within
$5\times10^{-5}$ over the full range, without interpolating a fitted curve.
The initial interval of width $10^{-12}$ above $e$ is enclosed
analytically: Bessel's inequality gives
$|\Tr(\rho_*U)|^2\ge1-(\varepsilon-e)/(q-e)$, and
$\lambda_{\min}(\rho_*)=p/[6(1+p)]$ converts this into
$D\le e+2.48\times10^{-5}$.
The common endpoint is attained by the Letter's exact optimum $U_{\alpha_*}$.
The accompanying data and checker record all endpoint witnesses,
coefficient covers and interval enclosures.

\section{Recovery with retained records and prescribed Kraus budgets}
\label{sec:sm-recorded}

Let $A_{si}:\mathbb C^d\to\mathbb C^{m_s}$ describe the instrument
$\cN(X)=\bigoplus_s\sum_i A_{si}XA_{si}^\dagger$, with
$\sum_{s,i}A_{si}^\dagger A_{si}=I_d$. The decoder receives outcome
$s$ and the corresponding quantum output. Its branch $\cR_s$ maps
$\mathbb C^{m_s}$ back to $\mathbb C^d$. We hold the instrument fixed.
A trace-preserving branch with at most $k_s$ Kraus operators exists exactly
when $k_sd\ge m_s$, since stacking its Kraus operators gives an
isometry from $\mathbb C^{m_s}$ into $\mathbb C^{k_sd}$ and splitting any
such isometry into $d$-row blocks gives a branch decoder.
Write $F_{\boldsymbol k}^*$ and $D_{\boldsymbol k}^*$ for the best squared
worst-input fidelity and the smallest normalized diamond error at these caps. Let
$\mathcal T_{\boldsymbol k}$ be the set of positive block matrices
$\tau=\bigoplus_s\tau_s$ of trace one with $\operatorname{rank}\tau_s\le k_s$.

\subsection{Optimal value and deterministic construction}

For $\tau\in\mathcal T_{\boldsymbol k}$, factor
$\tau_s=L_s L_s^\dagger$ with $k_s$ columns, padding with zeros. Define
\begin{equation}
 B_{s\ell}=\sum_i\overline{(L_s)_{i\ell}}A_{si},\qquad
 X_s(\rho)=\operatorname{col}_{\ell}(\rho B_{s\ell}^\dagger),\qquad
 a(\tau)=\min_\rho\phi_\tau(\rho),\quad
 \phi_\tau(\rho)=\sum_s\norm{X_s(\rho)}_1 .
 \label{eq:sm-recorded-source}
\end{equation}
Here $\operatorname{col}$ stacks matrices vertically and $\norm{\cdot}_1$
is the trace norm. The matrix $X_s^\dagger X_s$ equals
$G_s(\rho,\tau_s)=\sum_{i,j}(\tau_s)_{ji}A_{si}\rho^2A_{sj}^\dagger$.
Thus the objective depends only on $\tau$ and is convex in the common
density matrix $\rho$. We will prove
\begin{equation}
 F_{\boldsymbol k}^*=1-D_{\boldsymbol k}^*
 =\max_{\tau\in\mathcal T_{\boldsymbol k}}a(\tau)^2.
 \label{eq:sm-recorded-optimum}
\end{equation}

For an isometric stack $V_s=\operatorname{col}_\ell(R_{s\ell})$ of branch
Kraus operators, put $T=\sum_{s,\ell}R_{s\ell}B_{s\ell}$. For any density
matrix $\rho$,
\begin{equation}
 \smin(T)\le\norm{T\rho}_1\le\phi_\tau(\rho).
 \label{eq:sm-recorded-upper}
\end{equation}
The first inequality follows from $T^\dagger T\succeq\smin(T)^2 I$.
For the second, maximize $\operatorname{Re}\Tr(UT\rho)$ over unitaries
$U$. Each $U R_{s\ell}$ again stacks to an isometry, and its trace
pairing with $X_s$ is bounded by $\norm{X_s}_1$.
The coefficients of $T$ in the physical roundtrip Kraus list have total
squared norm one, so $\norm{T}\le1$. At each $\rho$ the trace-norm
pairing is attained by isometric polar completions, giving
$0\le\phi_\tau(\rho)\le1$.

For attainment, minimize $\phi_\tau$ on $\rho\succeq\eta I/d$ with
$0<\eta<1$, and call the minimizer $\rho_\eta$ and value $a_\eta$.
Complete the polar partial isometry of $X_s(\rho_\eta)$ to an isometry
$V_s$ with the same $k_sd$ rows. At faithful $\rho$ the right kernel of
$X_s(\rho)$ is the fixed space $\bigcap_\ell\ker B_{s\ell}^\dagger$.
On its orthogonal complement $X_s^\dagger X_s$ is positive definite,
so the trace norm is differentiable there. The polar-completion
columns on the kernel give zero in the derivative, because every
$X_s(H)$ annihilates that space. Thus for Hermitian $H$,
\[
 d\phi_\tau(\rho_\eta)[H]
 =\operatorname{Re}\Tr(HT_\eta),\qquad
 a_\eta=\operatorname{Re}\Tr(\rho_\eta T_\eta).
\]
Writing $\rho_\eta=\eta I/d+(1-\eta)\sigma$ in the first-order condition
shows that $\sigma$ minimizes the expectation of $\operatorname{Re}T_\eta$.
Since $\norm{T_\eta}\le1$, it follows that
\[
 \lambda_{\min}(\operatorname{Re}T_\eta)
 \ge\frac{a_\eta-\eta}{1-\eta},\qquad
 a(\tau)\le a_\eta\le(1-\eta)a(\tau)+\eta.
\]
The last inequality follows by mixing a minimizing state with $I/d$.
Taking a convergent subsequence of the isometric stacks as
$\eta\downarrow0$ yields a trace-preserving decoder within the same
budgets with $\smin(T)=a(\tau)$, by Eq.~\eqref{eq:sm-recorded-upper}.
The compactness argument applies to singular minimizing sources.

One common inverse polar unitary changes the coherent amplitude
$T$ to $|T|$ without increasing any branch rank. A
normalized coherent Kraus amplitude $B\succeq bI$ can be selected as
one Kraus operator by a Kraus rotation. Its trace against every
density is at least $b$. Subtracting $bI$ from that Kraus operator
gives the centered Gram $(1+b^2)I-2bB\preceq(1-b^2)I$. The centered
diamond bound in the Letter therefore gives $F\ge b^2$ and $D\le1-b^2$.
Applying this with $b=a(\tau)$ proves achievability.

For the converse, take a decoder of worst-input fidelity $f>0$, a
minimizing state $\rho_0$, and set
$\alpha_{s\ell i}=\Tr(\rho_0 R_{s\ell}A_{si})$ and
$(L_s)_{i\ell}=\alpha_{s\ell i}/\sqrt f$. The reference $\tau_s=L_sL_s^\dagger$
lies within the same rank caps. First-order optimality of $\rho_0$
gives $\operatorname{Re}T\succeq\sqrt f I$. Equation
\eqref{eq:sm-recorded-upper} implies $a(\tau)\ge\sqrt f$.
The case $f=0$ gives the same upper bound on $f$ trivially.
The reference maximum of $a(\tau)^2$ is attained by compactness and
continuity. Combining the two directions with $D\ge1-F$ gives the
value formula and simultaneous deterministic attainment.
The rank-constrained fidelity duality of B\'eny and
Oreshkov~\cite{BenyOreshkov2011} concerns trace-nonincreasing recovery.
The isometric completion above is what enforces trace preservation
without adding a Kraus operator at the prescribed budgets. In particular,
for square branches the rank-one budget attains the unrestricted value
exactly when some unrestricted maximizing reference has rank at most
one on every branch.

\subsection{Complete optimizer families}

Put $q=F_{\boldsymbol k}^*>0$ and $t=\sqrt q$. The vector of roundtrip
trace amplitudes $\alpha(\rho)$ ranges over a compact convex set as
$\rho$ varies. Its minimum-norm vector is unique. Thus for a
fidelity optimizer, $\alpha/t$ defines a canonical reference, even
when the minimizing source is nonunique. Rotating the decoder Kraus
operators by a unitary $W_s$ sends $L_s$ to $L_sW_s^{\mathsf T}$, so
$\tau_s$ is intrinsic to the physical decoder. Conversely, any two
factors of the same $\tau_s$ with $k_s$ columns differ by such a unitary.

Fix a maximizing reference and its factor $L_s$. The corresponding
normalized amplitude is $T$ and the unnormalized amplitude is $M=tT$.
For a trace-preserving list, $\operatorname{Re}M\succeq qI$ implies
$F\ge q$ by Cauchy--Schwarz on the trace amplitudes. At a minimizing
source, equality in Cauchy--Schwarz forces $\alpha=tL$ in these Kraus
coordinates. Hence the test gives exactly the fidelity optimizers
with that canonical reference, by the first-order condition above.

For $0<q<1$, put $\epsilon=1-q$. For any operational optimizer the
fidelity is $q$, since $D\ge1-F$. Equation~\eqref{eq:sm-coherent} gives
an attained certificate
\[
 \left\|I-\sum_j\gamma_j K_j\right\|^2
       +\epsilon\norm{\gamma}_2^2\le\epsilon
\]
in the physical roundtrip Kraus list $K_j$. At a minimizing source
the left side is bounded below by
\[
 |1-\gamma\cdot\alpha|^2+\epsilon\norm{\gamma}_2^2
 =\epsilon+\epsilon\norm{\gamma-\bar\alpha}_2^2
        +|\alpha\cdot(\gamma-\bar\alpha)|^2 ,
\]
where the dot denotes contraction without conjugation.
Thus $\gamma=\bar\alpha$, giving the operational test below. Conversely,
that test supplies the certificate with the same coefficient. The two
conditions on a trace-preserving decoder at a maximizing reference are
\begin{equation}
 \begin{aligned}
 \text{fidelity optimal:}\quad &\operatorname{Re}M\succeq qI,\\
 \text{operationally optimal:}\quad &\norm{I-M}\le1-q .
 \end{aligned}
 \label{eq:sm-recorded-tests}
\end{equation}
Ranging over all maximizing references and all admissible branch Kraus
lists gives every optimizer. At $q=1$ both conditions force $M=I$ by
$\norm{M}\le1$, and all optimal round trips are the identity. At $q=0$
all feasible decoders have $F=0$ and $D=1$.

For an optimal reference, choose any minimizing $\rho$ of
Eq.~\eqref{eq:sm-recorded-source}. A fidelity optimizer satisfies
$t\le\operatorname{Re}\Tr(\rho T)\le\phi_\tau(\rho)=t$, so every
branch saturates its trace-norm pairing. Its stack is the polar
partial isometry of $X_s(\rho)$ on the support of $X_s^\dagger X_s$,
and an isometric completion on the kernel. The two matrix tests
select the optimal completions. This gives the full physical families
after identifying Kraus lists that represent the same channel.

\subsection{Physical coordinates and the cost of completion}

The classification also has a form directly in Choi coordinates.
Use row vectorization, let $J_s$ be the physical branch Choi block,
and set $Q_s(\rho)=[\operatorname{vec}(\rho A_{si}^\dagger)]_i$.
Consider
\[
 Z_s=\begin{pmatrix}J_s&W_s\\ W_s^\dagger&K_s\end{pmatrix}\succeq0,\quad
 \Tr_{\rm out}J_s=I_{m_s},\quad
 \operatorname{rank}J_s\le k_s,\quad \sum_s\Tr K_s=q .
\]
Define $M=\sum_{s,i}\operatorname{unvec}(W_s[:,i])A_{si}$ and apply
Eq.~\eqref{eq:sm-recorded-tests}. Positivity of $Z_s$ gives for every source
\[
 f_\rho-2\operatorname{Re}\Tr(\rho M)+q
 =\sum_s\Tr\!\left[
   (Q_s^\dagger,-I)Z_s
   \begin{pmatrix}Q_s\\-I\end{pmatrix}\right]\ge0 .
\]
Here $f_\rho$ is the round-trip entanglement fidelity at $\rho$.
The fidelity inequality implies $F\ge q$. At a minimizing source
the sum is zero, so the columns of the stack of $Q_s$ over $-I$
lie in the kernel of $Z_s$.
Consequently $W_s=J_s Q_s$ and $K_s=Q_s^\dagger J_s Q_s$.
The lift has the same rank as $J_s$ and is unique. The operational
test implies the fidelity test and therefore selects exactly the
same operational optimizers as above. Conversely each optimizer
provides this lift. This formulation uses physical rank constraints
only. At the optimum $K_s/q$ is the canonical reference when $q>0$.

For a fixed optimal reference and $q>0$, use a minimal factor $L_s$
with $r_s=\operatorname{rank}\tau_s$ columns. In that coefficient
gauge write the decoder as a coherent part with operators $C_{s\ell}$
and a completely positive remainder $\mathcal Y_s$:
\[
 \cR_s(X)=\sum_{\ell=1}^{r_s}C_{s\ell}XC_{s\ell}^\dagger+\mathcal Y_s(X),
 \qquad S_s=\mathcal Y_s^\dagger(I_d)
       =I_{m_s}-\sum_{\ell=1}^{r_s}C_{s\ell}^\dagger C_{s\ell}.
\]
The matrix tests involve only the coherent operators. The remainder
can complete trace preservation within the budget exactly when
$S_s\succeq0$ and $\operatorname{rank}S_s\le d(k_s-r_s)$.
Indeed a square-root factor of $S_s$ can be divided into matrices
with $d$ rows, requiring exactly
$\lceil\operatorname{rank}S_s/d\rceil$ remainder Kraus operators.
For a nonzero block, the matrix of coherent columns is
$[\operatorname{vec} C_{s\ell}]_\ell
 =W_s L_s(L_s^\dagger L_s)^{-1}/\sqrt q$ by the canonical lift.
This determines the coherent part and the remainder uniquely in
the fixed factor coordinates. At a worst source the coherent
columns map under $Q_s^\dagger$ to $\sqrt q L_s$, whereas the
remainder's Choi range lies in $\ker Q_s^\dagger$. Their ranks
therefore add. The minimum total rank for the specified coherent
correction is $r_s+\lceil\operatorname{rank}S_s/d\rceil$.
A zero reference block has only the completion.

\section{Calibration after a stabilizer-code decoder}
\label{sec:sm-code}

To test output calibration after error correction, we use a fixed
deterministic decoder for the known
$[[8,2,3]]$ stabilizer code specified in Ref.~\cite{Hao2021}, within the
eight-qubit family of Ref.~\cite{Gottesman1996}. Six positive generators
are $XIZIYZXY$, $IXZZYXYI$, $IZXIYYZX$, $IZZYZXZZ$, $IIZIIIYX$ and
$ZZZZZZZZ$. The logical Pauli frame is
$X_1=XZIYIIII$, $X_2=ZYIIXIII$, $Z_1=ZZZIZIII$, $Z_2=ZIIZZZII$.
The code corrects every single-qubit error. All 64 syndromes receive a
minimum-weight Pauli recovery, with ties broken by site order and then
$X,Y,Z$. The primary comparison discards
the syndrome after decoding and permits one final logical unitary.

\begin{figure}[ht]
\includegraphics[width=\textwidth]{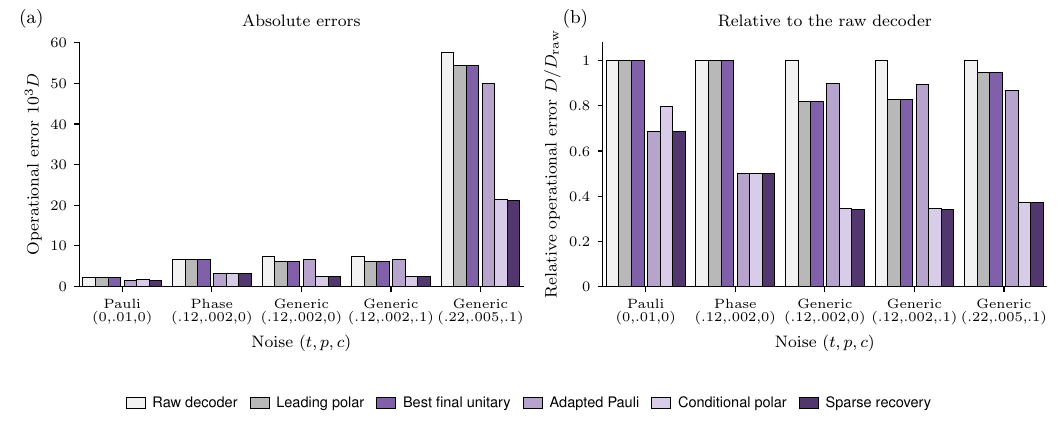}
\caption{Calibration after the deterministic $[[8,2,3]]$ decoder.
(a) Operational error $D$ for the five noise cases $(t,p,c)$.
(b) The same errors divided by the corresponding raw decoder error.
The first three bars discard the syndrome after the Pauli frame.
Adapted Pauli, conditional polar and sparse recovery retain it.
Sparse-recovery bars give certified upper bounds.}
\label{fig:code}
\end{figure}

Let $V$ encode the two logical qubits. The physical pulse is
\begin{equation}
 U_{t,c}=e^{-ict Z_0Z_1}\bigotimes_{i=0}^7e^{-it_i\bm n_i\cdot\bm\sigma_i},
 \qquad t_i=t[1+0.1((i\bmod3)-1)] .
\end{equation}
Generic unit axes repeat normalized $(1,1,1),(1,2,-1),(2,-1,1)$;
the phase control uses $(0,0,1)$. Independent physical $Z$ faults of
probability $p$ follow the pulse. For syndrome recovery $C_s$, the
decoded Kraus matrices are
\begin{equation}
 A_{s,e}=\sqrt{p^{|e|}(1-p)^{8-|e|}}\,V^\dagger C_sZ^eU_{t,c}V.
\end{equation}
Here $e\in\{0,1\}^8$ labels a fault pattern, $|e|$ is its Hamming
weight and $Z^e=\bigotimes_{i=0}^7 Z^{e_i}$.
The syndrome projection is implicit in $V^\dagger C_s$, whose support
is the corresponding syndrome subspace. Encoding and syndrome extraction
are ideal in this benchmark. Figure~\ref{fig:code} compares five noise
cases. Adapted Pauli chooses the logical Pauli with highest tracial
entanglement fidelity on each syndrome. Conditional polar takes the
inverse polar unitary of each syndrome's leading Kraus operator.
Sparse recovery uses a conditional unitary on all but at most three
syndromes, where it permits two Kraus operators.

Final-unitary correction reduces raw error by $5.3$--$17.9\%$ in the
generic cases; prior polar correction captures $99.90$--$99.98\%$ of
that gain. The SDP supplies a global final-unitary certificate, with
numerical $q/u$ gaps at most $1.29\times10^{-7}$. The Pauli and phase
controls have zero final-unitary gain. Conditional polar improves every
case further. In the Pauli control, adapted Pauli attains the optimum
with one Kraus operator per syndrome. In the three generic cases, the
sparse-recovery error bounds are about $0.8$--$1.1\%$ below the corresponding
conditional-polar errors.

\subsection{Sparse construction and resource certificate}
For the phase and generic cases, a trial decoder and its worst input supply the
environment reference by the construction in Sec.~\ref{sec:sm-recorded}.
We truncate each reference block to its selected rank and renormalize the
total trace. We then minimize the common-input objective
in Eq.~\eqref{eq:sm-recorded-source} and take the branch polar factors.
To certify a numerical decoder, we normalize its Kraus list on each branch
by right multiplication with $(\sum_\ell R_{s\ell}^\dagger R_{s\ell})^{-1/2}$.
We bound the effect of this normalization on the coherent amplitude and
apply Eq.~\eqref{eq:residual} after one common output polar correction.
This certifies the error of the constructed decoder within its rank caps.

For the generic case $(t,p,c)=(.12,.002,.1)$, let $D_*$ be the optimum over
all conditional CPTP decoders with the same syndrome record. We obtain
\begin{equation}
 0.0025067699<D_*\le D_{\rm sparse}<0.0025067886,
 \qquad D_{\rm sparse}-D_*<1.87\times10^{-8}.
 \label{eq:sm-code-sparse}
\end{equation}
Every conditional unital decoder has error above $0.0025075870$.
The sparse decoder uses rank two on syndromes $5,12,27$ and rank one
elsewhere, with the first listed stabilizer indexing the least significant
syndrome bit. At additive tolerance $2\times10^{-8}$ from $D_*$, rank two
on each of these three syndromes is necessary.

For the converse, use a fixed logical density matrix $\rho$ and row
vectorization. Define $C_s(\rho)=\sum_e\proj{\operatorname{vec}(\rho A_{s,e}^\dagger)}$.
If $J_s$ is the branch decoder's Choi matrix, its contribution to the
fidelity at $\rho$ is $\Tr(C_s(\rho)J_s)$. Hermitian matrices $Y_s,Z_s$
satisfying
\begin{equation}
 I_d\otimes Y_s+Z_s\otimes I_d\succeq C_s(\rho)
\end{equation}
bound the total fidelity at $\rho$ by $\sum_s\Tr(Y_s+Z_s)$ when
the branch maps are unital and trace preserving. Setting $Z_s=0$ on
a branch requires only trace preservation there. Thus setting all $Z_s$
to zero bounds unrestricted recovery, and allowing $Z_s$ only on one
branch bounds recovery constrained to be unital on that branch.
The inequality $D\ge1-F$ turns these fidelity bounds into error lower bounds.

We enclose the physical instrument and check the dual slacks and the
normalized sparse decoder's coherent bound using 192-bit outward
ball arithmetic. For each of syndromes $5,12,27$, the lower bound with
unitality on that syndrome alone exceeds the sparse upper bound by more
than $2\times10^{-8}$. A rank-one trace-preserving decoder on a square
branch is unitary and hence unital. Since $D_*$ is at most the sparse
upper bound, this excludes rank one on each of those branches at the
stated tolerance.

\section{Pure isometric targets and retained leakage}
\label{sec:sm-isometric}

Let $\Phi$ map a $d$-dimensional input to an $n$-dimensional output,
$n\ge d$, and let the ideal target be the
isometry $V$, $V^\dagger V=I$. Define
\begin{equation}
 F_V(\Phi)=\min_\rho\sum_j|\Tr(\rho V^\dagger A_j)|^2,
 \quad D_V(\Phi)=\tfrac12\norm{\Phi-\Ad_V}_\diamond.
\end{equation}
At a minimizing $\rho_0$, put $f=F_V(\Phi)$,
$\alpha_j=\Tr(\rho_0V^\dagger A_j)$ and
$T=\sum_j\bar\alpha_jA_j$. First-order optimality gives
$\operatorname{Re}(V^\dagger T)\succeq fI$, hence $\smin(T)\ge f$.
For $f>0$, the rectangular polar decomposition of $B=T/\sqrt f$ is
$B=WK$, where $W$ is an isometry and $K\succeq\sqrt fI$.
Extend $W$ and $V$ to output bases and choose an output unitary $U$
with $UW=V$.

The positive-amplitude norm bound extends directly. Suppose a Kraus
representation contains $A_1=WK$, where $W^\dagger W=I$ and
$K\succeq\mu I$ with $\mu\ge0$. Set $L_1=W(K-\mu I)$ and
$L_j=A_j$ for $j>1$. Then
\begin{equation}
 \begin{split}
 H&=(1+\mu^2)I-2\mu K,\quad 0\preceq H\preceq(1-\mu^2)I,\\
 (\Phi-\Ad_W)(X)
   &=\sum_jL_jXL_j^\dagger-\tfrac12W(HX+XH)W^\dagger.
 \end{split}
\end{equation}
Isometric embedding preserves trace norm, giving
$D_W(\Phi)\le1-\mu^2$. Consequently
\begin{equation}
 1-F_V(\Ad_U\Phi)\le D_V(\Ad_U\Phi)
       \le1-F_V(\Phi)\le D_V(\Phi).
\end{equation}
The zero-fidelity case uses the universal bound one. All unitary-closed
family consequences follow as in the main text. Optimizing the pure target gives
\begin{equation}
 \min_{W^\dagger W=I}D_W(\Phi)
 =1-\max_{W^\dagger W=I}F_W(\Phi)
 =1-\max_{\norm{c}_2=1}\smin\left(\sum_j\bar c_jA_j\right)^2.
\end{equation}
The same complementary-channel SDP and rank-one proof apply. The
extension covers leakage levels retained in the output when the final
correction controls that full space. A mixed-target example shows the
role of purity: the fixed output $I_2/2$ has squared fidelity $4/5$ and
distance $2/5$ from $\diag(9/10,1/10)$ under every output unitary.

\bibliography{references}